\documentclass[%
reprint,
superscriptaddress,
nofootinbib,
amsmath,amssymb, amsfonts,
showkeys,
prb
]{revtex4-2}

\usepackage{graphicx}% Include figure files
\usepackage{dcolumn}% Align table columns on decimal point
\usepackage{bm}% bold math
\usepackage{physics}
\usepackage[dvipsnames]{xcolor}
\usepackage[
    colorlinks,
    citecolor=RoyalBlue,
    linkcolor=RoyalBlue,
    urlcolor=RoyalBlue,
]{hyperref}
\usepackage{multirow}
\usepackage{mathtools}
\usepackage{verbatim}  % for comments
\usepackage[normalem]{ulem}
\newcommand{\iu}{\mathrm{i}\mkern1mu}
\newcommand{\eu}{\mathrm{e}\mkern1mu}

\begin{document}

% === MAIN TEXT START ===

% \title{Enhancement of excitonic radiative lifetime in van der Waals heterostructures integrated with resonant metasurfaces}
\title{Excitonic bound states in the continuum in \\ van der Waals heterostructure metasurfaces}

\author{Polina Pantyukhina}
\affiliation{
 Harbin Engineering University, Qingdao Innovation and Development Center, Qingdao, Shandong, China}
\affiliation{School of Physics and Engineering, ITMO University, St. Petersburg 197101, Russia}

\author{Andrey Bogdanov}
\email{a.bogdanov@hrbeu.edu.cn}
\affiliation{
 Harbin Engineering University, Qingdao Innovation and Development Center, Qingdao, Shandong, China}
\affiliation{School of Physics and Engineering, ITMO University, St. Petersburg 197101, Russia}

\author{Kirill Koshelev}%
\email{ki.koshelev@gmail.com}
\affiliation{University of New South Wales at Canberra, ACT 2600, Australia}%
\affiliation{Australian National University, Canberra ACT 2601, Australia}

\date{\today}

\begin{abstract}
We investigate the formation of {\it excitonic bound states in the continuum} in van der Waals (vdW) heterostructures composed of two-dimensional excitonic vdW layers and an optically resonant patterned vdW thin film. We show that the radiative losses of the exciton can be completely suppressed -- not through conventional methods such as total internal reflection, Bragg mirrors, or metallic layers -- but instead via destructive interference of exciton emission rates to distinct optical modes of the metasurface. We formulate the general conditions of excitonic BICs as a vanishing Purcell factor with non-vanishing vacuum local density of states at the exciton frequency. We propose a mechanism to achieve excitonic quasi-BICs with almost complete suppression of radiation via exciton coupling with a guided-mode resonance and multiple Fabry--P\'{e}rot modes. We show that in unpatterned vdW slabs, the Purcell factor suppression is defined exclusively by the slab's permittivity achieved via positioning the 2D exciton layer in the minimum of the mode electric field. We confirm through numerical simulations that, in periodically patterned heterostructure metasurfaces, the Purcell factor can be suppressed by more than five orders of magnitude, and this effect is not due to vanishing local electric fields. Our results demonstrate the formation of excitonic quasi-BICs and their potential for advancing quantum optics and information processing.
\end{abstract}


\maketitle

%%% SECTION 1
%%%%%%%%%%%%%%%%%%%%%%%%%%%%%%%%%%%%%%%
{\it Introduction}.---The strong excitonic response of semiconductor materials plays a pivotal role in advancing nanophotonics and nonlinear optics. Alongside excitons in semiconductor quantum wells, modern two-dimensional (2D) materials offer much stronger excitonic responses, enabling a wide range of novel optical phenomena. In particular, van der Waals (vdW) monolayers, such as transition metal dichalcogenides~\cite{manzeli20172d}, hexagonal boron nitride~\cite{cassabois2022exciton}, and topological insulators~\cite{xiang2021giant}, provide distinct advantages in photonics~\cite{lin2022engineering}, including large excitonic binding energies, high oscillator strengths, giant optical anisotropy~\cite{ermolaev2021giant,slavich2024exploring}, and tunable optical properties~\cite{wang2018colloquium}. These characteristics make vdW materials an ideal platform for sub-diffraction optical cavities, nanophotonic circuits, single-photon sources, and room-temperature polaritonic components~\cite{epstein2020highly}.

The integration of vdW monolayers with resonant metasurfaces significantly enhances light-matter interactions, enabling the realization of new photonic phenomena. The presence of Mie resonances and {\it bound states in the continuum} (BICs) in metasurface spectrum~\cite{koshelev2020dielectric} enables room-temperature boson condensation, polariton lasing, and ultra-high nonlinear responses~\cite{koshelev2018strong,verre2019transition,kravtsov2020nonlinear, ardizzone2022polariton,zograf2024combining,zograf2025ultrathin,ling2024nonlinear,tonkaev2024nonlinear}. The recent concept of {\it vdW heterostructure metasurfaces}~\cite{sortino2024van} provides a deeper integration of excitonic materials with resonant nanostructures, opening exciting avenues for ultrathin optical devices with atomic-scale precision~\cite{xu2025spatiotemporal,danielsen2025fourier}

%% FIGURE 1
%%%%%%%%%%%%%%%%%%%%%%%%%%%%%%%%%%%%%%%
\begin{figure}[t]
\includegraphics[width=0.98\linewidth]{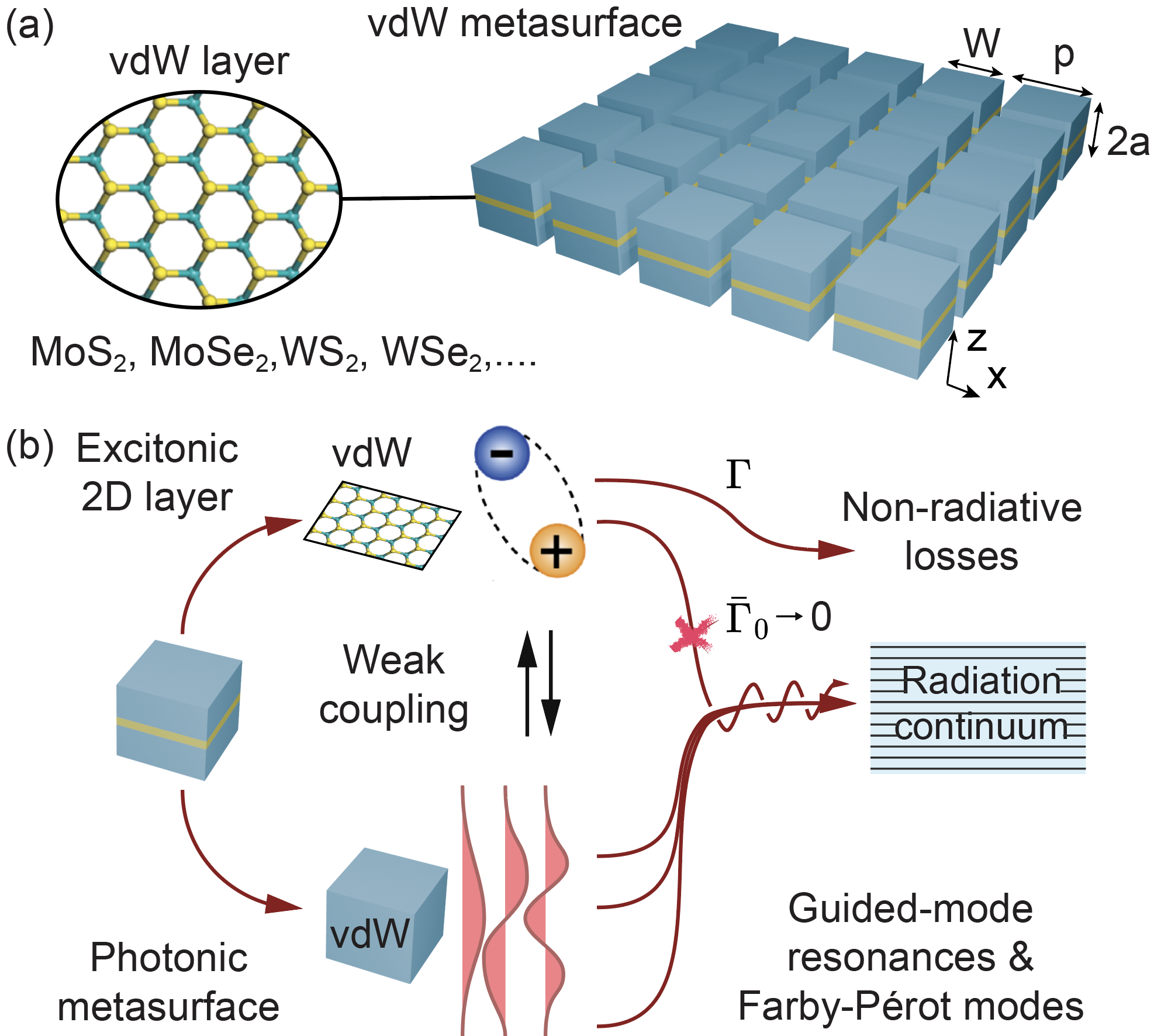}
\caption{\label{fig:1} {\bf Concept}. (a) Schematic of a van der Waals (vdW) heterostructure metasurface. (b) Schematic of excitonic BIC formation in the weak-coupling regime due to suppression of the spontaneous emission rate in the metasurface environment $\tilde{\Gamma}_0\to0$ via the destructive interference between the exciton coupling to guided-mode resonances and Fabry--P\'{e}rot modes.}
\end{figure}
%%%%%%%%%%%%%%%%%%%%%%%%%%%%%%%%%%%%%%%

While much of the recent research has focused on strong coupling between excitons and optical resonances, leading to the formation of polaritons~\cite{koshelev2018strong,lundt2019optical,lackner2021tunable,wu2024exciton,weber2023intrinsic,ding2025strong,schroder2025strong}, the weak coupling regime remains under-explored. In this regime, excitons interact with photonic modes without causing strong distortions to both the material and photonic subsystems, enabling precise control over excitonic energy and decay rates. This control is particularly valuable for fine-tuning device performance, where subtle adjustments to the radiative lifetime or spectral position can have a considerable impact. The changes in the spectral shift and radiative lifetime in this weak-coupling regime are well described by the theory of Lamb shift and Purcell factor~\cite{novotny2012principles,bayer2001inhibition}. Recently, 2D vdW monolayers integrated with resonant photonic structures, featuring complex local photonic density of states, were shown to achieve higher levels of Purcell enhancement~\cite{lee2017single}.

Although much of the current work has been devoted to enhancing the exciton emission rate, the reverse effect -- the suppression of the exciton spontaneous emission rate -- has received far less attention yet offers considerable promise. Increase of the exciton radiative lifetime is crucial for controlling excitonic dynamics in applications such as optical switches~\cite{feng2021all}, modulators~\cite{lynch2025full}, and sensors~\cite{runowski2024supersensitive}. Recent studies have demonstrated the suppression of the exciton spontaneous emission rate in systems with metallic mirrors and dielectric slabs~\cite{bayer2001inhibition, fang2019control}, however, a comprehensive theoretical framework for this effect is largely unexplored.

In this Letter, we investigate the formation of {\it excitonic states in the continuum} (BICs) in vdW heterostructure metasurfaces [see Fig.~\ref{fig:1}(a)]. 
% Excitonic BICs arise from the interplay between resonant photonic modes and excitonic states, where destructive interference completely suppresses exciton radiation losses. 
We first formulate the general conditions of excitonic BIC requiring a vanishing Purcell factor ${\rm F}$ with non-vanishing vacuum LDOS at the exciton frequency. We next show that ${\rm F}=0$ can be asymptotically achieved in weakly periodic structures in the vicinity of a guided-mode resonance frequency due to destructive interference of spontaneous emission through multiple channels corresponding to Fabry--P\'{e}rot modes [see Fig.~\ref{fig:1}(b)]. We show that in unpatterned vdW slabs, the maximal suppression of ${\rm F}$ is achieved via the conventional mechanism of positioning the vdW layer in the minimum of a mode electric field. We then numerically show that for periodically patterned heterostructure metasurfaces, the suppression of ${\rm F}$ reaches more than five orders of magnitude, indicating formation of excitonic quasi-BICs. We finally discuss the differences between photonic and excitonic BICs.

%%% SECTION 2
%%%%%%%%%%%%%%%%%%%%%%%%%%%%%%%%%%%%%%%
%\section{\label{sec:level3}Homogeneous dielectric slab}

{\it Excitonic BIC concept}.---In this section, we define excitonic BICs in periodic heterostructure metasurfaces, schematically shown in Fig.~\ref{fig:1}(a). The eigenmode spectrum of hybrid exciton-photon states can be found by diagonalizing the full Hamiltonian composed of quantized excitonic states of a periodically patterned 2D vdW layer, photonic modes of periodic vdW film metasurface and their light-matter interaction governed by the coupling matrix $\hat{V}$ (see Sec.~S1.A~\cite{SM}).  Assuming harmonic time dependence for the fields $\eu^{-\iu\omega t}$, we can write the matrix elements of $V_{\nu,n}$ between $\nu$-th quantized excitonic state and $n$-th optical quasi-normal mode (QNM) as~\cite{gerace2007quantum}
\begin{equation}    
\begin{aligned}
&V_{\nu,n}(z_0)=\sqrt{2c\Gamma_\nu}\sum_\mathbf{g}\Phi^*_{\nu,\mathbf{g}}\mathbf{ e}_{\nu}\cdot\mathbf{E}_{n,\mathbf{g}}(z_0),\\
&V^\dagger_{\nu,n}(z_0)=\sqrt{2c\Gamma_\nu}\sum_\mathbf{g}\Phi_{\nu,\mathbf{g}}\mathbf{ e}^*_{\nu}\cdot\mathbf{E}_{n,-\mathbf{g}}(z_0).
\end{aligned}
\label{eq:m1}
\end{equation}
Here, $\Phi_{\nu}(x,y)$ and $\mathbf{E}_{n}(\mathbf{r})$ is the center-of-mass envelope of exciton wavefunction and the QNM electric field, respectively, $\Gamma_\nu$ is the vacuum spontaneous emission rate of $\nu$-th exciton in a patterned vdW layer at normal incidence, $\mathbf{e}_{\nu}$ is the exciton polarization vector. The vector subscript in Eqs.~(\ref{eq:m1}) means the in-plane Fourier transform to the reciprocal space with vector $\mathbf{g}$. Here, we assumed $\Phi(z)\propto\delta(z-z_0)$, where $z_0$ is the vdW layer vertical position. The full Hamiltonian of the system represents a generalized Hopfield matrix with $V_{\nu,n}$ in the upper and $V^\dagger_{\nu,n}$ in the lower triangle, respectively (see Sec.~S1.B~\cite{SM}). The hybrid exciton-photon BIC can be formally defined as an eigenstate of the diagonalized system with energy above the light line and zero radiative losses.

In case of weak exciton-photon coupling, we can define the excitonic BIC as an exciton state of the heterostructure metasurface characterized with zero spontaneous emission rate and at least one open radiation channel at the frequency $\omega_{\rm BIC}$. For the fundamental quantized 2D exciton with frequency $\omega_0$ and vacuum spontaneous emission rate $\Gamma_0$, its spontaneous emission rate $\tilde{\Gamma}_0$ in the metasurface environment can be written via the Purcell factor ${\rm F}(\omega_0)$~(see Sec.~S1.C~\cite{SM})
\begin{equation}
\tilde{\Gamma}_0(\omega_0) ={\rm F}(\omega_0)\pi\rho(\omega_0),
\label{eq:m2}
\end{equation}
where $\rho(\omega)=1/\pi$ is the one-dimensional LDOS of vacuum radiation states~\cite{ochiai2001nearly}. Equation~\eqref{eq:m2} shows that the BIC condition, generalized to a finite Bloch in-plane vector $\mathbf{k}_{\rm b}$, can be written in the sub-diffraction regime as
\begin{equation}
\begin{cases}
\rm F({\mathbf{k}_{\rm B}};\omega_{\rm BIC})=0,\\
\rho(\mathbf{k}_{\rm B};\omega_{\rm BIC})\ne 0.
\end{cases}
\label{eq:m3}
\end{equation}
Equation~\eqref{eq:m3} is analogous to the condition of photonic BIC, that requires vanishing coupling amplitudes to all open radiation channels with a non-vanishing LDOS.

% We note periodic boundary conditions discretize the radiation continuum, which allows to achieve photonic BICs in subwavelength dielectric metasurfaces~\cite{koshelev2023bound}. Consequently, we show below that small number of open radiation channels allows for excitonic BICs in subwavelength heterostructure metasurfaces.

{\it Theoretical framework for Purcell factor}.--- 
The Purcell factor of the fundamental quantized exciton with frequency $\omega_0$ in a heterostructure metasurface at normal incidence can be expressed as (see Sec.~S1.C~\cite{SM})
\begin{equation}
{\rm F}(\omega_0)=-2\sum_{\mathbf{g},\mathbf{g}^\prime}\operatorname{Im}\left[\Phi^*_{\mathbf{g}}\mathbf{e}\cdot\hat{\mathbf{G}}_{\mathbf{g},\mathbf{g}^\prime}( z_0,z_0;\omega_0)\cdot\mathbf{e}^*\Phi_{\mathbf{g}^\prime}\right],
\label{eq:m4}
\end{equation}
where $\hat{\mathbf{G}}_{\mathbf{g},\mathbf{g}^\prime}(z,z^\prime;\omega)$ is the in-plane Fourier transform to the reciprocal space of  the dimensionless dyadic electric-electric Green function (GF). We expand $\hat{\mathbf{G}}_{\mathbf{g},\mathbf{g}^\prime}( z_0,z_0;\omega_0)$ for $|z_0|<a$ into the pole series at complex QNM frequencies $\omega_n-\iu\gamma_n$ using the Mittag-Leffler theorem $\hat{\mathbf{G}}_{\mathbf{g},\mathbf{g}^\prime}(z,z^\prime;\omega)=\sum_{n} {c \mathbf{E}_{n,\mathbf{g}}(z)\otimes\mathbf{E}_{n,-\mathbf{g}^\prime}(z^\prime)}/{(\omega-\omega_n+\iu\gamma_n)}$~\cite{neale2020resonant}. We then substitute the pole expansion into Eq.~\eqref{eq:m4} and obtain the QNM series for the Purcell factor (see Sec.~S1.D~\cite{SM})
\begin{equation}
{\rm F}(z_0;\omega_0)=-\sum_{n} \operatorname{Im}\left[\frac{V_{n}(z_0)V_{n}^\dagger(z_0)}{\Gamma_0(\omega_0-\omega_n+\iu\gamma_n)}\right],
\label{eq:m5}
\end{equation}
where the index $\nu$ is omitted. We note that Eq.~\eqref{eq:m5} extends the rigorous approach for evaluating the Purcell factor in photonic structures weakly coupled to quantum emitters via the QNM expansion to periodic heterostructure systems~\cite{sauvan2013theory,muljarov2016exact}. 

For $\omega_0$ away from the diffraction threshold and in the vicinity of an isolated guided-mode resonance (GMR) with a frequency $\omega_n-\iu \gamma_n$, the contribution of Rayleigh anomalies~\cite{karavaev2025emergence} and other GMRs to ${\rm F}(\omega_0)$ can be neglected. The resonant GMR contribution can be separated in Eq.~\eqref{eq:m5} from the remaining contribution of Fabry--P\'{e}rot (FP) modes, and written in the form of generalized Fano formula~\cite{bogdanov2019bound}
\begin{equation}
{\rm F}(z_0;\omega_0)={\rm F}^{(\rm wg)}-{\rm F}^{(\rm env)}+\frac{\left(q\gamma_n+\omega_0-\omega_{n}\right)^2}{(\omega_0-\omega_{n})^2+\gamma_n^2}{\rm F}^{(\rm env)},
\label{eq:m6}
\end{equation}
where ${\rm F}^{(\rm wg)}(z_0;\omega_0)$ is the contribution of the FP modes, $q(z_0)=-\cot{(\arg{[V_{n}(z_0)V^\dagger_{n}(z_0)]}/2)}$ is the Fano asymmetry parameter, and  ${\rm F}^{(\rm env)}(z_0)={\left|V_{n}(z_0)V^\dagger_{n}(z_0)\right|}/{[\gamma_n\Gamma_0(1+q^2)]}$ is the smooth envelope. The minimal value of ${\rm F}(\omega_0)$ in Eq.~\eqref{eq:m6} is achieved at $\omega_0=\omega_{n}-q\gamma_n$ and is equal to ${\rm F}^{(\rm wg)}-{\rm F}^{(\rm env)}$. Therefore, the excitonic BIC condition can be written as
\begin{equation}
\begin{aligned}
& \omega_{\rm BIC}=\omega_n-q_n(z_0)\gamma_n,\\
&{\rm F}^{(\rm wg)}(z_0;\omega_{\rm BIC})={\rm F}^{(\rm env)}(z_0).
\end{aligned}    
\label{eq:m7}
\end{equation}
%% FIGURE 2
%%%%%%%%%%%%%%%%%%%%%%%%%%%%%%%%%%%%%%%
\begin{figure}[t]
\includegraphics[width=0.9\linewidth]{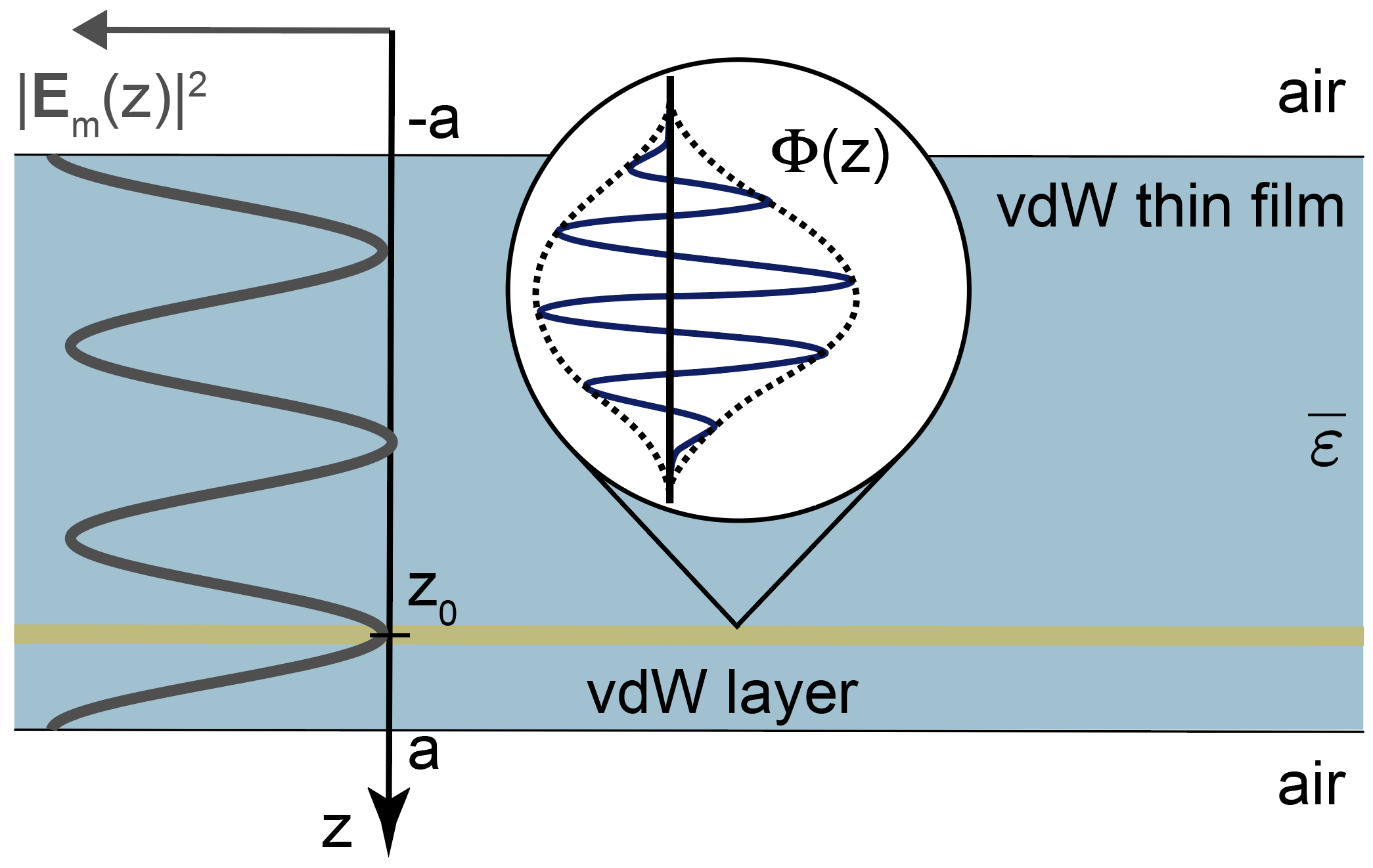}
\caption{{\bf Geometry of the heterstructure slab}. Electric field intensity $|\mathbf{E}_m(z)|^2$ for the FP QNM with $m=3$ and exciton center-of-mass envelope wavefunction $\Phi(z)$ in the $z$-direction, shown with gray and blue solid lines, respectively.  
\label{fig:2}}
\end{figure}
%%%%%%%%%%%%%%%%%%%%%%%%%%%%%%%%%%%%%%%
The excitonic BIC condition in Eq.~\eqref{eq:m7} can be fulfilled in the asymptotic limit of a weakly periodically modulated metasurface with $\delta S=(1-W^2/p^2)\ll1$, where $W$ is the meta-atom side length, and $p$ is the metasurface period (see Sec.~S1.D.2~\cite{SM}). As an example, we consider the $s$-polarized GMR at normal incidence, $\mathbf{E}_{n}=\mathbf{e}_y{E}_{n}$. For $\delta S\ll1$, $\Phi_{\mathbf{g}}\simeq\delta_{\mathbf{g},\mathbf{0}}$ and the coupling amplitudes in Eq.~\eqref{eq:m1} for $y$-polarized excitons become $V_{n}\simeq V^\dagger_{n}\simeq\sqrt{2c\Gamma_0}\left<{E}_{n}(\mathbf{r})\right>$. The in-plane average of the electric field $\left<{E}_{n}\right>\simeq{\rm F}^{(\rm wg)}\delta S\alpha_n$ describes the weak coupling of the GMR to the open s-polarized radiation channel via the discrete spectrum of leaky FP modes, where $\alpha_n$ is an auxiliary complex-valued function. By reciprocity, the GMR linewidth is $\gamma_n\simeq 2c{\rm F}^{(\rm wg)}(\delta S\operatorname{Im}[\alpha_n])^2$. Similarly, the GMR frequency $\omega_n$ deviates from the frequency of the doubly degenerate guided modes on the scale of $(\delta S)^2$. The Fano parameter becomes $q=-\operatorname{Re}[V_{n}]/\operatorname{Im}[V_{n}]\to-\operatorname{Re}[\alpha_{n}]/\operatorname{Im}[\alpha_{n}]$ for $\delta S\to 0$. Combined, the first BIC condition in Eq.~\eqref{eq:m7} becomes asymptotically fulfilled. At the same time, the envelope in Eq.~\eqref{eq:m6} transforms to ${\rm F}^{(\rm env)}={(\operatorname{Im}[V_{n}])^2}/{(\gamma_n\Gamma_0)}\to {\rm F}^{(\rm wg)}$ and the second BIC condition in Eq.~\eqref{eq:m7} is also asymptotically fulfilled.

{\it Effective heterostructure slab}.---  We next analyze the contribution of FP modes to the Purcell factor of the effective heterostructure slab  ${\rm F}^{(\rm wg)}$ composed of vdW thin film of the permittivity $\overline{\varepsilon}$ and 2D vdW layer, shown schematically in Fig.~\ref{fig:2} (see also Sec.~S2.A~\cite{SM}). As an example, at normal incidence for $s$-polarization the transversal component of the photonic GF is~\cite{muljarov2011brillouin} 
\begin{equation}
{\rm G}(\tilde z_0,\tilde z_0;\tilde\omega_0)=\frac{1+r^2\eu^{2\iu\pi\tilde\omega_0  }+2r\eu^{\iu\pi\tilde\omega_0 }\cos{(\pi\tilde\omega_0\tilde z_0)}}{2\iu \sqrt{\overline{\varepsilon}}\left(1-r^2\eu^{2\iu\pi\tilde\omega_0  }\right)}.
\label{eq:m8}
\end{equation}
Here, $r=(\sqrt{\overline{\varepsilon}}-1)/(\sqrt{\overline{\varepsilon}}+1)$ is the Fresnel reflection coefficient, and  $\tilde z_0=z_0/a$ and $\tilde\omega_0=\omega_0/\omega_{\rm ph}$, where $\omega_{\rm ph} = \pi c/(2\sqrt{\overline{\varepsilon}}a)$. Substituting Eq.~\eqref{eq:m8} into Eq.~\eqref{eq:m4}, we obtain for linearly polarized excitons
\begin{equation}
{\rm F}^{(\rm wg)}= (1-r)^2\frac{\left[1+r^{2}  +2r\cos(\pi\tilde\omega_0)\cos(\pi\tilde\omega_0\tilde z_0)\right]}{\left[(1+r^{2})^2  -4r^2\cos^2(\pi\tilde\omega_0)\right]}.
\label{eq:m9}
\end{equation}

%% FIGURE 3
%%%%%%%%%%%%%%%%%%%%%%%%%%%%%%%%%%%%%%%
\begin{figure*}[t]
\includegraphics[width=0.9\linewidth]{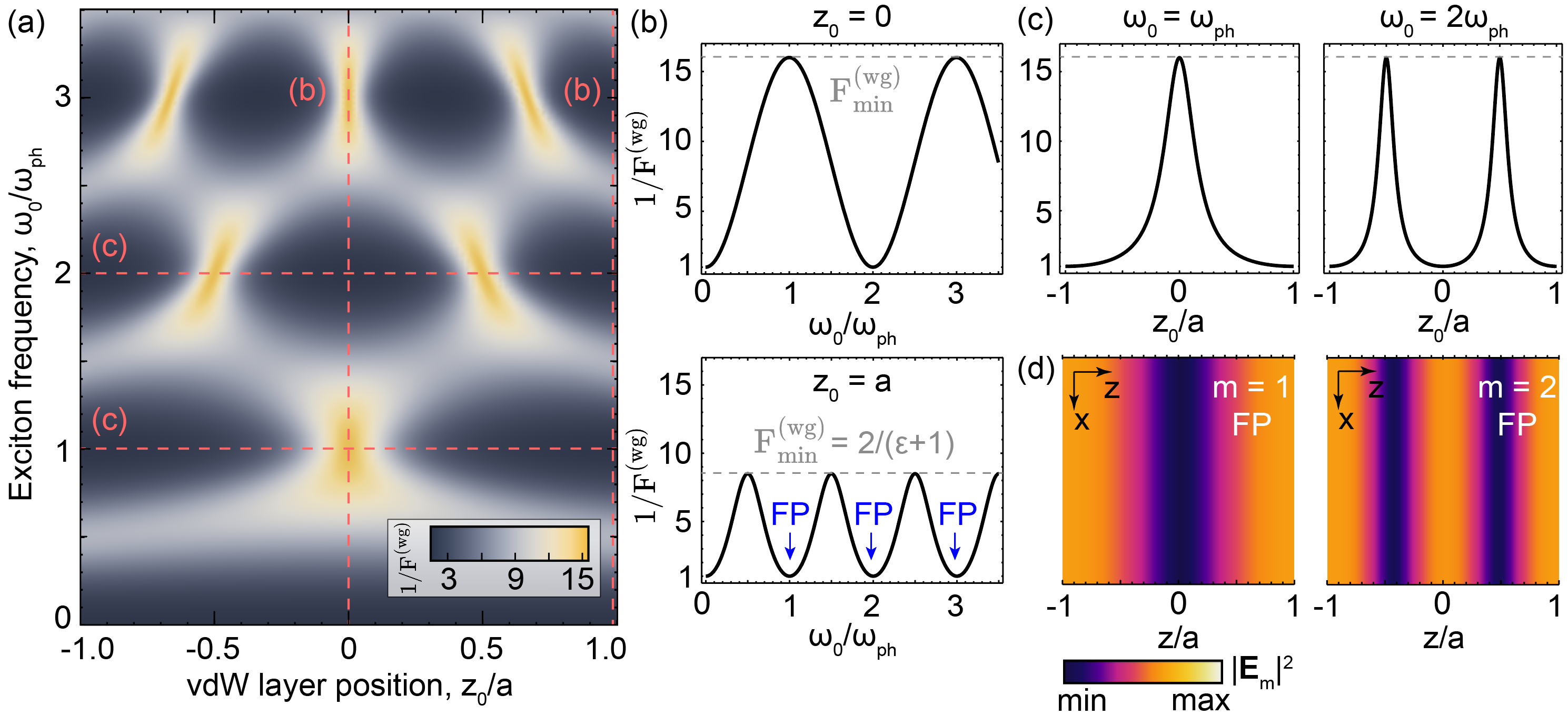}
\caption{\label{fig:3} {\bf Purcell factor in effective heterostructure slab}. (a) Inverse Purcell factor $1/{\rm F}^{(\rm wg)}$ [Eq.~\eqref{eq:m9}] vs. normalized 2D layer position $\tilde z_0$ and exciton frequency $\tilde\omega_0$ for $\overline{\varepsilon}=16$. Pink dashed lines correspond to panels (b,c) [see Eq.~\eqref{eq:m10} for $|z_0|<a$]. (b) $1/{\rm F}^{(\rm wg)}$ vs. $\tilde\omega_0$ for $z_0=0$ (upper panel) and $z_0=a$ (lower panel). (c) $1/{\rm F}^{(\rm wg)}$ vs. $\tilde z_0$ for resonant $ \tilde\omega_0=1,2$. The fundamental limit ${\rm F}^{(\rm wg)}_\text{min}$ in (b,c) is shown with gray dashed lines. (d) The electric field profiles $|\mathbf{E}_{m}|^2$ of FP modes for $m=1,2$. }
\end{figure*}
%%%%%%%%%%%%%%%%%%%%%%%%%%%%%%%%%%%%%%%

The extrema of the function ${\rm F}^{(\rm wg)}(\tilde z_0;\tilde\omega_0)$ within the slab volume $|\tilde z_0| <1$ are given by simultaneous solutions of $\sin(\pi\tilde\omega_0)=0$ and $\sin(\pi\tilde\omega_0\tilde z_0)=0$ that are satisfied at 
\begin{equation}
\begin{aligned}
&\tilde\omega_0=m,\quad m=1,2,\ldots\\
&\tilde z_0=l/m,\quad l=-(m-1),\ldots (m-1).
\end{aligned}
\label{eq:m10}
\end{equation}
The minimum and maximum values of the Purcell factor are ${\rm F}^{(\rm wg)}_{\rm min}=1/\overline{\varepsilon}$ and ${\rm F}^{(\rm wg)}_{\rm max}=1$, respectively. The meaning of the conditions in Eq.~\eqref{eq:m10} becomes clear from the analysis of the FP QNMs of the slab~\cite{muljarov2011brillouin, lalanne2018light, sauvan2022normalization}. For $\overline{\varepsilon} \gg 1$, the FP frequencies $\omega_{m} - \iu \gamma_m$ with $\omega_m > 0$ are given by $\omega_{m} = m \omega_{\rm ph}$, and $\gamma_m=\gamma\simeq c/(\overline{\varepsilon} a)$~\cite{muljarov2011brillouin}. Thus, $\omega_{\rm ph} = \omega_{m+1} - \omega_m$ is the FP frequency spacing, and $\tilde\omega_0=m$ is the resonant condition for the FP modes. The magnitude of the FP electric field within the slab $|\tilde z|\le 1$ is $|\mathbf{E}_{m}(\tilde z)|^2\simeq\cos^2[\pi m(\tilde z-1)/2]/(2\overline{\varepsilon} a)$. Consequently, $\tilde z_0=l/m$ is the condition of the $l$-th extremum of $|\mathbf{E}_{m}(\tilde z)|^2$, see also Fig.~\ref{fig:2} for $m=3$. 
% We also note that at oblique incidence at an angle $\theta$, ${\rm F}^{(\rm wg)}_\text{max}(\theta)$ remains $1$, while ${\rm F}^{(\rm wg)}_\text{min}(\theta)=\cos^2\theta/(\overline{\varepsilon}-\sin^2\theta)$. At $\theta\to \pi/2$ ${\rm F}^{(\rm wg)}_\text{min}(\theta)\to 0$, which indicates the transformation of leaky FP modes into bound guided modes.

Figure~\ref{fig:3}(a) shows the dependence of the inverse Purcell factor $1/{\rm F}^{(\rm wg)}$ on $\tilde z_0$ and $\tilde\omega_0$ via Eq.~\eqref{eq:m9} for $\overline{\varepsilon}=16$ at normal incidence. The map cross-sections at a constant $\tilde z_0$ and $\tilde\omega_0$ are shown in Figs.~\ref{fig:3}(b,c), respectively. The calculated ${\rm F}^{(\rm wg)}_\text{min}$ and ${\rm F}^{(\rm wg)}_\text{max}$ are shown with gray dashed lines. Figures~\ref{fig:3}(c) and \ref{fig:3}(d) show the matching of $1/{\rm F}^{(\rm wg)}$ at $\tilde\omega_0=m$ and $|\mathbf{E}_{m}(\tilde{z})|^2$ for $m=1,2$. 

The number of FP modes that contribute to ${\rm F}^{(\rm wg)}$ on- and off- resonance varies depending on the ratio of $V_m$ to the exciton-photon frequency mismatch, $\eta_m=|V_m|/\sqrt{(\tilde\omega_0-m)^2\omega_{\rm ph}^2+\gamma^2}$ (see Sec.~S2.B~\cite{SM}), where we assume small exciton non-radiative loss rate $\Gamma\ll\gamma$. In the resonant regime, $\eta_{m}$ of the resonant QNM dominates, thus a single FP is sufficient to describe ${\rm F}_{\rm max}^{(\rm wg)}=1$. In the off-resonant regime $\eta_{m}$ of neighboring QNMs are comparable, requiring many of them to describe ${\rm F}_{\rm min}^{(\rm wg)}=1/\overline{\varepsilon}$.

%% FIGURE 4
%%%%%%%%%%%%%%%%%%%%%%%%%%%%%%%%%%%%%%%
\begin{figure*}[t]
\includegraphics[width=0.95\linewidth]{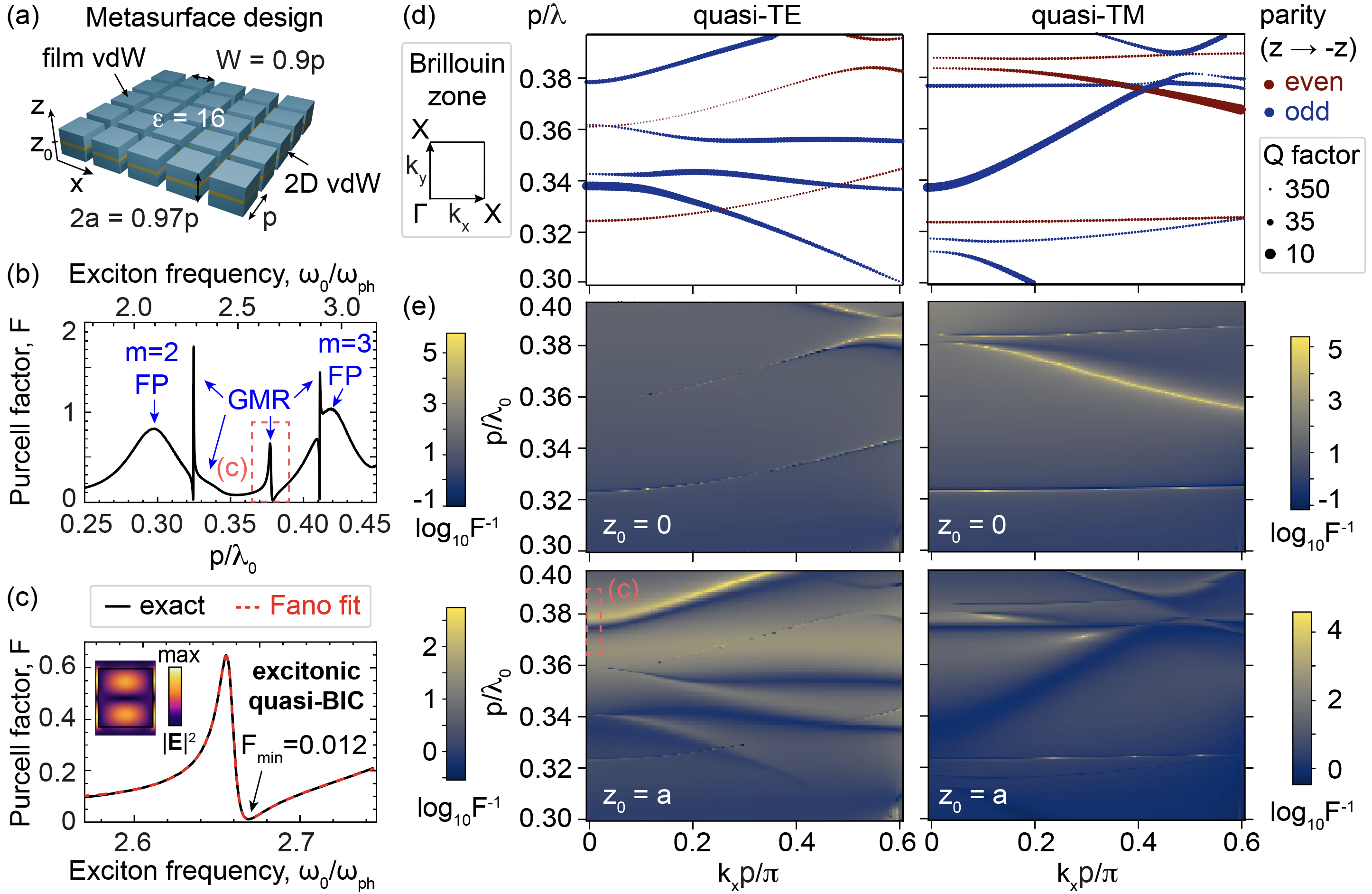}% Here is how to import EPS art
\caption{\label{fig:4} {\bf Purcell factor in patterned heterostructure metasurface}. (a) Schematic of a heterostructure metasurface. The 2D vdW layer is positioned at $z=z_0$. (b) Purcell factor ${\rm F}$ of TE-polarized excitons vs. $p/\lambda_0$ and $\omega_0/\omega_{\rm ph}$ at normal incidence for $z_0 = a$. The frequencies of excited GMRs and FP modes are shown with blue arrows. (c) Comparison of exact (black) and fitted to the Fano lineshape (dashed red) ${\rm F}$ in the vicinity of a GMR frequency $\tilde{\omega}_0=2.66$. Inset: the GMR electric field profile cross section in the $zy$ plane. (d) Dispersion diagram $p/\lambda$ vs. $k_xp/\pi$ for quasi-TE- and quasi-TM-polarized GMRs of the metasurface along the $\Gamma-$X direction ($k_y=0$). Inset on the left shows the Brillouin zone. Red and blue color shows even and odd parity with respect to $z\to -z$ mirror symmetry, respectively. The line thickness is inversely proportional to the mode Q factor. (e) Logarithmic scale of ${\rm F}^{-1}$ vs. $p/\lambda_0$ (and $\omega_0/\omega_{\rm ph}$) and  $k_xp/\pi$ for $z_0=0$ (top) and $z_0=a$ (bottom) along the $\Gamma-$X direction ($k_y=0$) in the case of TE (left) and TM (right) exciton polarization.}
\end{figure*}
%%%%%%%%%%%%%%%%%%%%%%%%%%%%%%%%%%%%%%%

%%% SECTION 3
%%%%%%%%%%%%%%%%%%%%%%%%%%%%%%%%%%%%%%%
%\section{\label{sec:3}Resonant dielectric metasurface}

{\it vdW heterostructure metasurface}.--- With the knowledge of ${\rm F}^{(\rm wg)}(z_0;\omega_0)$, we return to the analysis of Eqs.~(\ref{eq:m6}) and (\ref{eq:m7}) for a heterostructure metasurface composed of a vdW thin film and a 2D vdW layer inside it or on its surface, shown schematically in Fig.~\ref{fig:4}(a). The metasurface is composed of square blocks of width $W$ and height $2a$ arranged into a square lattice of period $p$. 

We focus on the doubly degenerate pair of bright circularly polarized 1s exciton states of WSe$_2$ monolayer with the wavelength $\lambda_0=2\pi c/\omega_0\simeq730$~nm and the opposite projection of spin. In the vicinity of normal incidence, the QNM polarization is linear, thus, the metasurface environment mixes the degenerate exciton states into TE and TM linearly polarized ones. For the sake of simplicity, we neglect both the ellipticity of the exciton polarization and anisotropy of $\varepsilon$~\cite{munkhbat2022optical}, that are considerable only at large oblique angles of incidence. Consequently, we choose the permittivity of vdW thin film $\varepsilon\simeq 16$, that approximately corresponds to WS$_2$ at $730$~nm. 

Following the prediction of Eqs.~(\ref{eq:m6}) and (\ref{eq:m7}), we numerically analyze the spectrum of ${\rm F}$ for linearly polarized excitons at $\lambda_0\simeq730$~nm. In the model, we use the frequency-domain solver of COMSOL Multiphysics with Floquet periodic boundary conditions and unit cell parameters $2a=0.97p$, $W=0.9p$, where $p$ is varied. The 2D vdW layer is modeled as a periodically patterned directional surface current emitting at $730$~nm with the direction of current along TE or TM polarization. Consequently, ${\rm F}$ is evaluated as the ratio of the radiated power from the current within the heterostructure metasurface and vacuum environment, respectively. 

The FP mode frequencies in the effective slab model [see Eq.~\eqref{eq:m10}] are the integer number of $\omega_{\rm ph} = \pi c/(2\sqrt{\overline{\varepsilon}}a)$, where $\overline{\varepsilon}=1+(\varepsilon-1)W^2/p^2=0.19+0.81\varepsilon$~\cite{poddubny2013hyperbolic}. Figure~\ref{fig:4}(b) shows ${\rm F}(p/\lambda_0)$ for TE-polarized excitons at $z_0=a$, corresponding to the range $1.8<\tilde\omega_0<3.2$ between the second and third FP modes. The Purcell factor spectrum exhibits asymmetric Fano features in the vicinity of metasurface GMRs. We focus on the GMR at $\tilde{\omega}_0= 2.66$, with the mode profile shown in the inset of Fig.~\ref{fig:4}(c). The spectrum of ${\rm F}$ demonstrates a perfect agreement with the Fano profile in Eq.~\eqref{eq:m6} with the parameters $q=-2.1$, ${\rm F}^{\rm (env)}=0.12$, and the mode quality factor $240$.

Next, we study the suppression of ${\rm F}(p/\lambda_0)$ for a broader range of $p/\lambda_0$ and in-plane wavevectors along the $\Gamma$-X direction. Figure~\ref{fig:4}(d) shows the spectra of quasi-TE- and quasi-TM-polarized GMR with even and odd out-of-plane symmetry. Figure~\ref{fig:4}(e) shows the corresponding $1/{\rm F}$ in the log-scale demonstrating suppression up to five orders of magnitude depending on $z_0$. 

The observed suppression of Purcell factor substantially exceeds $1/{\rm F}^{\rm (wg)}_{\rm min}=\overline{\varepsilon}$, manifesting the formation of {\it excitonic quasi-BICs}. We can interpret them as exciton states decoupled from the radiation continuum due to the destructive interference of its coupling rates to a GMR and multiple FP modes, required to describe the contribution of ${\rm F}^{\rm (wg)}$ correctly. Unlike conventional optical state density suppression (e.g., via total internal reflection or Bragg mirrors), the local electric field is not minimal in the vicinity of Purcell factor minimum (see Sec.~S3.A~\cite{SM}). 

We also note that excitonic quasi-BIC can be realized even in low-contrast heterostructure metasurfaces, as there are no restrictions on $\varepsilon$ in Eq.~\eqref{eq:m7}. We show numerically that ${\rm F}$ can be suppressed up to eight orders of magnitude for $\varepsilon=4$ and $z_0=0$ (see Sec.~S3.B~\cite{SM}).

{\it Discussion}.--- We next compare conventional photonic BICs and the proposed excitonic BICs. Firstly, Friedrich-Wintgen (accidental) photonic BICs have nontrivial far-field polarization properties around the selected k-space point where radiation contributions vanish. In contrast, excitons lack intrinsic far-field polarization, instead, their topological features can be analyzed through phase divergence in the phase of the coupling coefficients~\cite{bulgakov2017topological,bezus2018bound} [see Eq.~\eqref{eq:m1}]. Secondly, radiative quality factor of photonic BICs exhibits a power law  divergence in k-space~\cite{zhang2025super}. For excitonic BICs, the radiative Q factor behavior in k-space can deviate considerably due to the effect of exciton-photon energy mismatches on the weak coupling conditions (see Fig.~S2~\cite{SM}). Lastly, Figs.~\ref{fig:4}(d,e) show that excitonic quasi-BICs do not form near the photonic BICs wavelengths. In this regime, the Purcell factor is dominated by the uncompensated FP contribution ${\rm F}^{\rm (wg)}$.

{\it Conclusions}.--- We have investigated excitonic bound states in the continuum (BICs) in van der Waals metasurfaces composed of a patterned vdW thin film and 2D vdW layer. We have formulated the general conditions for excitonic BICs and proposed to achieve them via the exciton coupling to multiple quasi-normal modes of the vdW film.  We have shown that near a guided-mode resonance  frequency, the Purcell factor exhibits the asymmetric Fano lineshape, with the minimal value contributed by the guided-mode resonance and multiple broadband Fabry--P\'{e}rot  modes.
The numerical results confirm the formation of excitonic quasi-BICs via the suppression of Purcell factor by more than $5$ orders of magnitude even for low-index vdW films. Our work opens new possibilities for flexible control of excitonic optical properties through photonic design in vdW heterostructure metasurfaces, with potential applications in quantum information processing.

\begingroup
\renewcommand{\addcontentsline}[3]{}
\begin{acknowledgments}
The authors acknowledge useful discussions with Ivan Iorsh and Mikhail Glazov at early stages of the project. K.K acknowledges Ivan Toftul for proofreading the initial version of the manuscript.
\end{acknowledgments}
\endgroup

\begingroup
\renewcommand{\addcontentsline}[3]{}
\bibliography{bibliography}
\endgroup

\clearpage
\onecolumngrid

\begin{center}
  \textbf{\large Supplementary Material}
\end{center}

% Optional: reset figure and table counters
\setcounter{figure}{0}
\setcounter{table}{0}
\setcounter{equation}{0}
\renewcommand{\thefigure}{S\arabic{figure}}
\renewcommand{\thetable}{S\arabic{table}}
\renewcommand{\thesection}{S\arabic{section}}
\renewcommand{\theequation}{S\arabic{equation}}

\title{\SM\\
Excitonic bound states in the continuum \\ in van der Waals heterostructure metasurfaces}

\author{Polina Pantyukhina}
\affiliation{
 Harbin Engineering University, Qingdao Innovation and Development Center, Qingdao, Shandong, China}
\affiliation{School of Physics and Engineering, ITMO University, St. Petersburg 197101, Russia}

\author{Andrey Bogdanov}
\email{a.bogdanov@hrbeu.edu.cn}
\affiliation{
 Harbin Engineering University, Qingdao Innovation and Development Center, Qingdao, Shandong, China}
\affiliation{School of Physics and Engineering, ITMO University, St. Petersburg 197101, Russia}

\author{Kirill Koshelev}%
\email{ki.koshelev@gmail.com}
\affiliation{University of New South Wales at Canberra, ACT 2600, Australia}%
\affiliation{Australian National University, Canberra ACT 2601, Australia}

\maketitle
\begin{quote}\textbf{Abstract:} Section ~S1A contains the derivation of general model for light-matter coupling between 2D excitons and electromagnetic modes in patterned heterostructure metasurfaces. Section ~S1B shows how the problem can be formulated via a non-Hermitian Hamiltonian with a generalized non-symmetric Hopfield matrix. Section~S1C contains the analysis of the Purcell factor and derivation of the condition of excitonic bound states in the continuum (BICs). Section~S1D contains the derivation of excitonic BICs frequency in the weak perturbation regime and general case. Section~S2A provides detailed analysis of Purcell factor in effective heterostructure slabs with a 2D vdW layer at oblique incidence. Section~S2B contains the evaluation of the effective number of interfering Fabry-Perot modes. Section~S3A  describes the relation between the near-field enhancement and the far-field Purcell factor spectrum. Section~S3.B provides additional calculations for a low-index heterostructure metasurface with the permittivity $4$.
\end{quote}

\tableofcontents

\maketitle

\section{Exciton and exciton-polariton BICs in heterostructure metasurfaces}

\begin{figure}[t]
\includegraphics[width=0.3\linewidth]{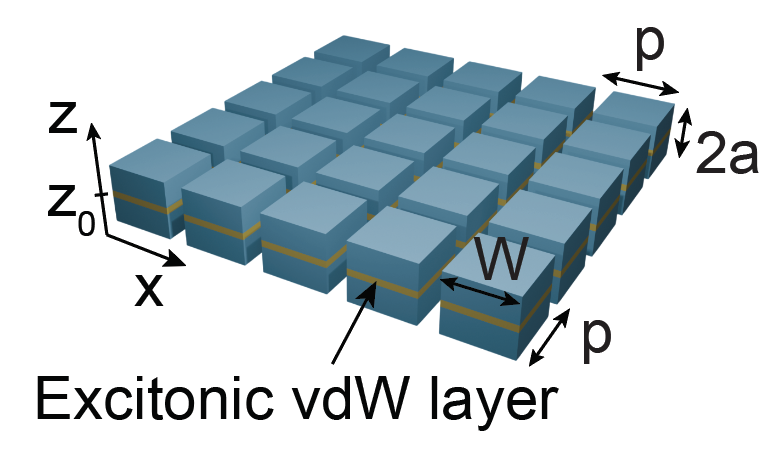}
\caption{\label{fig:00} Schematic of a heterostructure metasurface of thickness $2a$, period $p$, and square meta-atom width and length $W$.}
\end{figure}

In this section, we find the condition for exciton and exciton-polariton BICs in periodically patterned 2D vdW layers in heterostructure metasurfaces, schematically shown in Fig.~\ref{fig:00}. The metasurface thickness is $2a$, period is $p$, and square meta-atom width and length is $W$. We consider normal incidence and zero in-plane Bloch wavevector components. 

\subsection{Derivation of polariton eigenmode equation for heterostructure metasurface}

Light-matter interaction in a heterostructure metasurface  couples excitons of a vdW layer and electromagnetic modes of the dielectric metasurface, both periodically patterned in the $x-y$ plane. The dynamics of a coupled system is described by the many-body 2D exciton wavefunction $\ket{\Phi(t)}$ and electric field $\mathbf{E}(\mathbf{r},t)$. The nonlocal polarization $\mathbf{P}^{\rm (exc)}(\mathbf{r}, t)$ of the 2D vdW layer generated by the excited exciton states can be written as
\begin{equation}
4\pi\mathbf{P}^{\rm (exc)}(\mathbf{r},t)=\frac{8}{\pi a^2_{\rm B}}\bra{\delta(z-z_0)\Psi(t)}\hat{\mathbf{d}}\ket{\delta(z-z_0)\Psi(t)}+{\rm c.c.},  
\label{eq:1001}
\end{equation}
where $a_{\rm B}$ is the 2D Bohr radius, $z_0$ is the vertical position of the 2D vdW layer, $\hat{\mathbf{d}}$ is the operator of dipole moment. Assuming that the 2D exciton density is much smaller than the saturation density, we can separate the many- body wavefunction into the ground state $\ket{0}$ and weak perturbation due to excited exciton states as $\ket{\Psi(t)}=\ket{0}+\Delta\Phi(x,y,t)\ket{\varphi(\mathbf{\rho},z_0,z_0,t)}$, where $\ket{\varphi(\mathbf{\rho},z_0,z_0)}$ and $\Delta\Phi(x,y,t)(x,y,t)$  are the electron-hole and center-of-mass envelope wave functions, respectively, $\mathbf{\rho}$ is the relative in-plane distance between hole and electron. We assume that the excitons have the same polarization $\mathbf{e}$, thus we can expand $\bra{\varphi}\hat{\mathbf{d}}\ket{0}={\mathbf{d}}={d}\mathbf{e}$, where $d=-\iu a_{\rm B} \sqrt{\pi\hbar c\Gamma_0/(4\omega_0)}$~\cite{ivchenko2005optical}, where $\Gamma_0$ and $\omega_0$ are the spontaneous emission rate and frequency, respectively, of a non-patterned 2D vdW layer in the spectral range of interest.

Assuming harmonic dependence of fields in time $\eu^{-\iu\omega t}$, we can write Eq.~\eqref{eq:1001} transformed to the Fourier domain as~\cite{ivchenko2005optical} 
\begin{equation}
4\pi\mathbf{P}^{\rm (exc)}(\mathbf{r};\omega)=\frac{8}{\pi a^2_{\rm B}}\delta(z-z_0)\left[\mathbf{d}^*\Delta\Phi(x,y;\omega)+\mathbf{d}\Delta\Phi^*(x,y;-\omega)\right],
\label{eq:101}
\end{equation}
where we used $\bra{0}\hat{\mathbf{d}}\ket{0}=0$. Assuming normal incidence, we can transform all functions of $x,y$ to the reciprocal space, e.g.,
\begin{equation}
\begin{aligned}
&\mathbf{E}(\mathbf{r};\omega)=\sum_{\mathbf{g}}\eu^{\iu (g_x x+g_y y)}\mathbf{E}_{\mathbf{g}}(z;\omega),\\
&\Delta\Phi(x,y;\omega)=\sum_{\mathbf{g}}\eu^{\iu (g_x x+g_y y)}\Delta\Phi_{\mathbf{g}}(\omega).
\end{aligned}
\label{eq:1}
\end{equation}
Here, $\mathbf{g}=[g_{x},g_{x}]$ are reciprocal lattice vectors.
%% FIGURE S00
%%%%%%%%%%%%%%%%%%%%%%%%%%%%%%%%%%%%%%%
%%%%%%%%%%%%%%%%%%%%%%%%%%%%%%%%%%%%%%%

We can write the coupled Schr\"odinger and Maxwell equations in the reciprocal space within the first-order time-dependent perturbation theory as~\cite{ivchenko2005optical}
\begin{equation}
\begin{aligned}
&\sum_{\mathbf{g}^\prime}\hat{\mathcal{L}}_{\mathbf{g},\mathbf{g}^\prime}^{\rm (exc)}(\omega)\Delta\Phi_{\mathbf{g}^\prime}(\omega)=-\mathbf{d}\cdot\mathbf{E}_{\mathbf{g}}(z_0;\omega),\\
&\sum_{\mathbf{g}^\prime}\hat{\mathbf{L}}^{\rm (ph)}_{\mathbf{g},\mathbf{g}^\prime}(z;\omega)\mathbf{E}_{\mathbf{g}^\prime}(z;\omega)=-4\pi \frac{\omega^2}{c^2}\mathbf{P}^{\rm (exc)}_{\mathbf{g}}(z;\omega).
\end{aligned}    
\label{eq:2}
\end{equation} 
Here,  the Schr\"odinger $\hat{\mathcal{L}}^{\rm (exc)}(x,y;\omega)$ and Maxwell $\hat{\mathbf{L}}^{\rm (ph)}(\mathbf{r};\omega)$ operators in Eq.~\eqref{eq:2} are given by
\begin{equation}
\begin{aligned}
&\hat{\mathcal{L}}_{\mathbf{g},\mathbf{g}^\prime}^{\rm (exc)}(\omega)=\delta_{\mathbf{g}, \mathbf{g}^{\prime }}\left(\hbar\omega-\hbar\omega_0 - \frac{\hbar^2\mathbf{g}^2}{2m_{\rm (exc)}}\right)-\hat{\mathcal{U}}_{\mathbf{g}-\mathbf{g}^{\prime }},\\
&\hat{\mathbf{L}}^{\rm (ph)}_{\mathbf{g},\mathbf{g}^\prime}(z;\omega)=\hat{\mathbf{I}}\frac{\omega^2}{c^2}\varepsilon_{\mathbf{g}-\mathbf{g}^{\prime}}(z)-\delta_{\mathbf{g}, \mathbf{g}^{\prime}}\nabla_\mathbf{g}\times\nabla_\mathbf{g}\times,
\end{aligned}   
\label{eq:3}
\end{equation}
where $\varepsilon(\mathbf{r})$ is the periodic permittivity function, $m_{\rm (exc)}$ is the exciton effective mass, and $\hat{\mathcal{U}}(x,y)$ is the quantum potential formed by periodical patterning of 2D vdW layer with air gaps, and $\nabla_\mathbf{g}=[\iu g_x\ \ \iu g_y\ \  \partial_z]^\mathsf{T}$.

The exciton center-of-mass envelope wavefunction and electric field in Eq.~\eqref{eq:2} can be separated as
\begin{equation}
\Delta\Phi_{\mathbf{g}}=\frac{8\omega}{\pi ca^2_{\rm B}}\sum_{\mathbf{g}^\prime,\mathbf{g}^{\prime\prime}}\hat{\mathcal{G}}^{\rm (exc)}_{\mathbf{g},\mathbf{g}^\prime}(\omega) \mathbf{d}\cdot\hat{\mathbf{G}}^{\rm (ph)}_{\mathbf{g}^\prime,\mathbf{g}^{\prime\prime}}(z_0,z_0;\omega)\cdot\left[\mathbf{d}^*\Delta\Phi_{\mathbf{g}^{\prime\prime}}(\omega)+\mathbf{d}\Delta\Phi^*_{\mathbf{g}^{\prime\prime}}(-\omega)\right],
\label{eq:4}
\end{equation}    
where $2a$ is the metasurface thickness. The exciton and photonic Green functions (GFs) in Eq.~\eqref{eq:4} are defined as solutions of 
\begin{equation}
\begin{aligned}
&\sum_{\mathbf{g}^{\prime\prime}}\hat{\mathcal{L}}_{\mathbf{g},\mathbf{g}^{\prime\prime}}^{\rm (exc)}(\omega)\hat{\mathcal{G}}_{\mathbf{g}^{\prime\prime},\mathbf{g}^\prime}^{\rm (exc)}(\omega)=\delta_{\mathbf{g},\mathbf{g}^\prime},\\
&\sum_{\mathbf{g}^{\prime\prime}}\hat{\mathbf{L}}^{\rm (ph)}_{\mathbf{g},\mathbf{g}^{\prime\prime}}(z;\omega)\hat{\mathbf{G}}^{\rm (ph)}_{\mathbf{g}^{\prime\prime},\mathbf{g}^\prime}(z,z^\prime;\omega)=\frac{\omega}{c}\hat{\mathbf{I}}\delta_{\mathbf{g},\mathbf{g}^\prime}\delta(z-z^\prime).
\end{aligned}  
\label{eq:5}
\end{equation}

The exciton center-of-mass envelope wavefunction $\Delta\Phi_{\nu}(x,y)$ can be written as a sum of excited exciton states $\Phi_{\nu}(x,y)$ with amplitudes $c_\nu(\omega)$,
\begin{equation}
\Delta\Phi_{\mathbf{g}}=\sum_\nu c_\nu(\omega)\Phi_{\nu,\mathbf{g}}
\label{eq:6}
\end{equation}
where $c_\nu(\omega)$ are amplitudes, and $\nu$ is the index of excited exciton states that includes polarization and in-plane quantization indices due to periodic patterning. The states $\Phi_{\nu,\mathbf{g}}$ in Eq.~\eqref{eq:6} are defined as solutions of eigenvalue problem
\begin{equation}
\sum_{\mathbf{g}^{\prime}}\hat{\mathcal{L}}_{\mathbf{g},\mathbf{g}^{\prime}}^{\rm (exc)}(\omega_\nu)\Phi_{\nu,\mathbf{g}^\prime}=0,
\label{eq:7}
\end{equation}
where $\omega_\nu$ are real-valued positive eigenfrequencies. The excitonic GF $\hat{\mathcal{G}}^{\rm (exc)}(\omega)$ can be also expanded into basis eigenstates as
\begin{equation}
\hat{\mathcal{G}}_{\mathbf{g},\mathbf{g}^{\prime}}^{\rm (exc)}(\omega)=\sum_\nu\frac{\Phi_{\nu,\mathbf{g}}\Phi^*_{\nu,\mathbf{g}^\prime}}{(\hbar\omega-\hbar\omega_\nu+\iu\hbar\Gamma)},
\label{eq:8}
\end{equation}
where $\Gamma\ll\omega_\nu$ is a small non-radiative decay rate of excitons added to the exciton poles phenomenologically.

We can substitute Eqs.~(\ref{eq:6},\ \ref{eq:8}) into  Eq.~\eqref{eq:4}, and obtain a set of algebraic equations for wavefunction amplitudes with frequency-dependent coefficients
\begin{equation}
\begin{aligned}
&\sum_\mu\left\{\delta_{\mu,\nu}(\omega_\nu-\iu\Gamma)c_\mu(\omega)+\frac{8\omega}{c \pi \hbar a^2_{\rm B}}\sum_{\mathbf{g},\mathbf{g}^\prime}\Phi^*_{\nu,\mathbf{g}}{\mathbf{d}}\cdot\hat{\mathbf{G}}^{\rm (ph)}_{\mathbf{g},\mathbf{g}^\prime}(z_0,z_0;\omega)\cdot\left[{\mathbf{d}}^*\Phi_{\mu,\mathbf{g}^\prime} c_\mu(\omega)+{\mathbf{d}}\Phi^*_{\mu,\mathbf{g}^\prime}c^*_\mu(-\omega)\right]\right\}=\omega c_\nu(\omega),\\
&\sum_\mu\left\{\delta_{\mu,\nu}(-\omega_\nu-\iu\Gamma)c^*_\mu(-\omega)-\frac{8\omega}{c \pi \hbar a^2_{\rm B}}\sum_{\mathbf{g},\mathbf{g}^\prime}\Phi_{\nu,\mathbf{g}}{\mathbf{d}}^*\cdot\hat{\mathbf{G}}^{\rm (ph)}_{\mathbf{g},\mathbf{g}^\prime}(z_0,z_0;\omega)\cdot\left[{\mathbf{d}}\Phi^*_{\mu,\mathbf{g}^\prime} c^*_\mu(-\omega)+{\mathbf{d}}^*\Phi_{\mu,\mathbf{g}^\prime}c_\mu(\omega)\right]\right\}=\omega c^*_\nu(-\omega),
\end{aligned}
\label{eq:9}
\end{equation}
where we used the property $\hat{\mathbf{G}}^{\rm (ph)}_{\mathbf{g},\mathbf{g}^\prime}(z_0,z_0;-\omega)=-\left[\hat{\mathbf{G}}^{\rm (ph)}_{\mathbf{g},\mathbf{g}^\prime}(z_0,z_0;\omega)\right]^*$. For the most general case with excited states $\nu$ characterized with different dipole moments  $\mathbf{d}_{\nu}=d_{\nu}\mathbf{e}_{\nu}$. We can also introduce auxiliary excitonic states with an index $\bar{\nu}$ defined as $\omega_{\bar{\nu}}=-\omega_{\nu}$, $\Phi_{\bar{\nu},\mathbf{g}}=\Phi^*_{\nu,\mathbf{g}}$,  ${\mathbf{d}}_{\bar{\nu}}={\mathbf{d}}_\nu^*$, and amplitudes $c_{\bar{\nu}}(\omega)=c^*_\nu(-\omega)$. The resulting set of equations for $c_{\tilde{\nu}}(\omega)$, where $\tilde{\nu}={\bar{\nu},\nu}$ is the combined exciton index, reads
\begin{equation}
\sum_{\tilde{\mu}}\left[\delta_{\tilde{\mu},\tilde{\nu}}(\omega_{\tilde{\nu}}-\iu\Gamma)+\frac{8\omega}{c \pi \hbar a^2_{\rm B}}\sum_{\mathbf{g},\mathbf{g}^\prime}\Phi^*_{\tilde{\nu},\mathbf{g}}{\mathbf{d}}_{\tilde{\nu}}\cdot\hat{\mathbf{G}}^{\rm (ph)}_{\mathbf{g},\mathbf{g}^\prime}(z_0,z_0;\omega)\cdot{\mathbf{d}}_{\tilde{\nu}}^*\Phi_{\tilde{\mu},\mathbf{g}^\prime}\right]c_{\tilde{\mu}}(\omega)=\omega c_{\tilde{\nu}}(\omega).   
\label{eq:10}
\end{equation}

\subsection{Pole expansion of the photonic Green function and Hamiltonian formulation}

The  photonic Green function in Eq.~\eqref{eq:10} can be decomposed into pole contributions for $|z|,|z^\prime|<a$ using the Mittag-Leffler theorem~\cite{neale2020resonant}
\begin{equation}
\hat{\mathbf{G}}^{\rm (ph)}_{\mathbf{g},\mathbf{g}^\prime}(z,z^\prime;\omega)=c\sum_{n} \frac{ \mathbf{E}_{n,\mathbf{g}}(z)\otimes\mathbf{E}_{n,-\mathbf{g}^\prime}(z^\prime)}{(\omega-\omega_n+\iu\gamma_n)}. 
\label{eq:11}
\end{equation}
Here, $\mathbf{E}_{n,\mathbf{g}}(z)$ are the electric fields of the electromagnetic quasi-normal modes (QNMs) of the dielectric metasurface in the reciprocal space with the complex-valued frequencies $ck_n=(\omega_n-\iu\gamma_n)$ satisfying 
\begin{equation}
\sum_{\mathbf{g}^\prime}\hat{\mathbf{L}}^{\rm (ph)}_{\mathbf{g},\mathbf{g}^\prime}(z;k_n)\mathbf{E}_{n,\mathbf{g}^\prime}(z)=0,
\label{eq:12}
\end{equation}
where $n$ labels the QNM modes and includes the discretized contribution of the continuum of cut modes at the diffraction threshold frequencies~\cite{neale2020resonant}.

We can substitute Eq.~\eqref{eq:11} into  Eq.~\eqref{eq:10} and get
\begin{equation}
\sum_{\tilde{\mu}}\left[\delta_{\tilde{\mu},\tilde{\nu}}(\omega_{\tilde{\nu}}-\iu\Gamma)+\frac{\omega}{\sqrt{\omega_{\tilde{\nu}}\omega_{\tilde{\mu}}}}\sum_n\frac{V_{\tilde{\nu},n}(z_0)V^\dagger_{\tilde{\mu},n}(z_0)}{(\omega-\omega_n+\iu\gamma_n)}\right]c_{\tilde{\mu}}(\omega)=\omega c_{\tilde{\nu}}(\omega),   
\label{eq:12}
\end{equation}
where $V_{\tilde{\nu},n}(z)$ and $V^\dagger_{\tilde{\nu},n}(z)$ are the exciton-photon coupling amplitudes. Using $d_{\tilde{\nu}}=-\iu a_{\rm B} \sqrt{\pi \hbar c\Gamma_{\tilde{\nu}}/(4\omega_{\tilde{\nu}})}$, the coupling amplitudes in Eq.~\eqref{eq:12} can be defined as
\begin{equation}
\begin{aligned}
&V_{\tilde{\nu},n}(z)\equiv\iu\sqrt{\frac{8\omega_{\tilde{\nu}}}{\pi \hbar a^2_{\rm B}}}\sum_\mathbf{g}\Phi^*_{\tilde{\nu},\mathbf{g}}\mathbf{d}_{\tilde{\nu}}\cdot\mathbf{E}_{n,\mathbf{g}}(z)=\sqrt{2c\Gamma_{\tilde{\nu}}}\sum_\mathbf{g}\Phi^*_{\tilde{\nu},\mathbf{g}}\mathbf{e}_{\tilde{\nu}}\cdot\mathbf{E}_{n,\mathbf{g}}(z),\\
&V^\dagger_{\tilde{\nu},n}(z)\equiv-\iu\sqrt{\frac{8\omega_{\tilde{\nu}}}{\pi\hbar a^2_{\rm B}}}\sum_\mathbf{g}\Phi_{\tilde{\nu},\mathbf{g}}\mathbf{d}^*_{\tilde{\nu}}\cdot\mathbf{E}_{n,-\mathbf{g}}(z)=\sqrt{2c\Gamma_{\tilde{\nu}}}\sum_\mathbf{g}\Phi_{\tilde{\nu},\mathbf{g}}\mathbf{e}^*_{\tilde{\nu}}\cdot\mathbf{E}_{n,-\mathbf{g}}(z).
\end{aligned}  
\label{eq:13}
\end{equation}

%% FIGURE 0
%%%%%%%%%%%%%%%%%%%%%%%%%%%%%%%%%%%%%%%
\begin{figure}[t]
\includegraphics[width=0.45\linewidth]{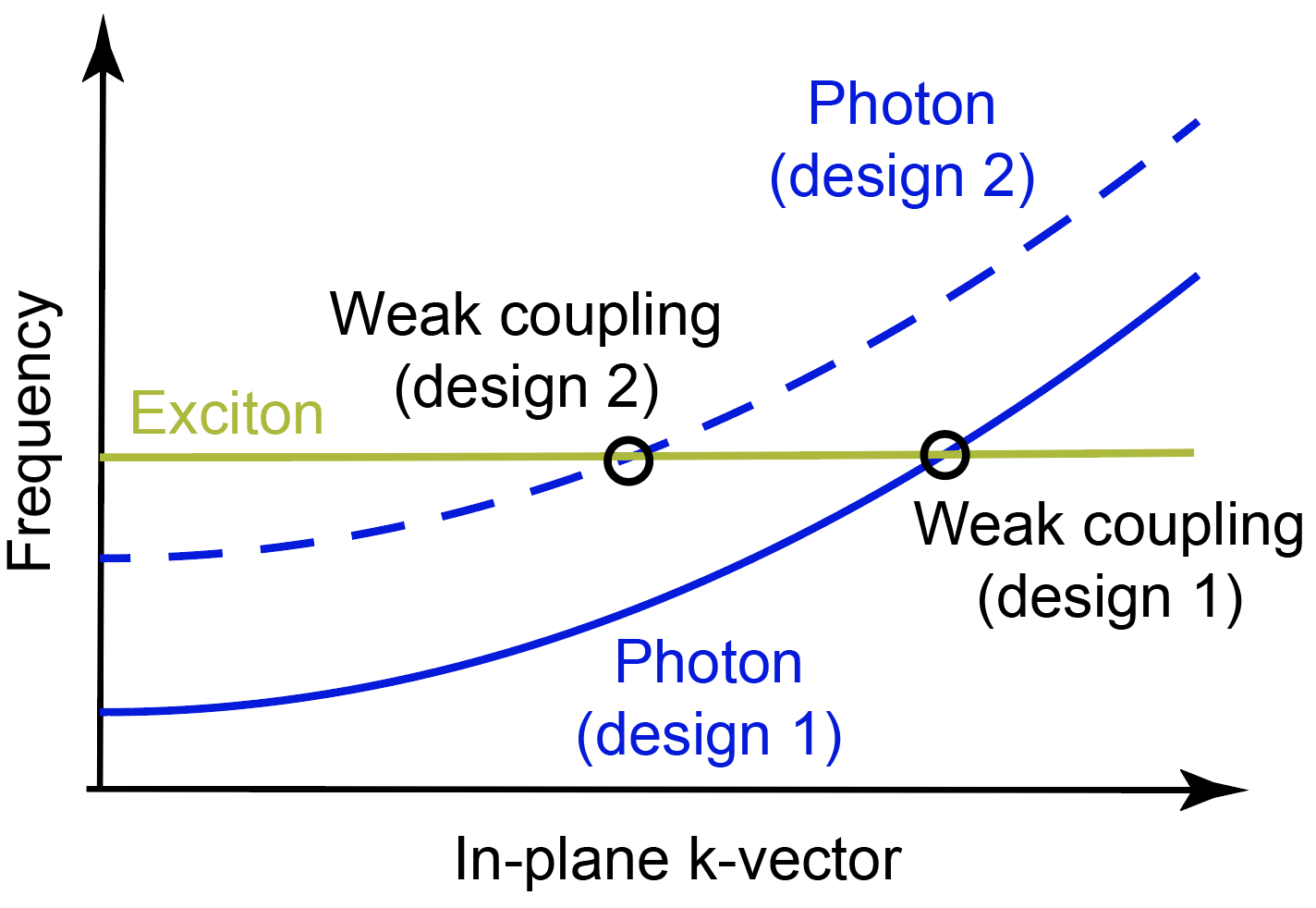}
\caption{{\bf Schematic of weak coupling}. Schematic of weak coupling dispersion diagram for exciton and photon. Two different designs of photonic structures are shown (solid and dashed blue line) leading to weak coupling regime at two different in-plane k-vector values. 
\label{fig:0}}
\end{figure}
%%%%%%%%%%%%%%%%%%%%%%%%%%%%%%%%%%%%%%%

A general equation for complex-valued frequencies of exciton-polaritons $\omega$ can be obtained via Eq.~\eqref{eq:12} in the form $\operatorname{det}|\hat{\mathcal{H}}-\hbar\omega\hat{\mathcal{I}}|=0$, where $\hat{\mathcal{I}}$ is the unitary operator, and $\hat{\mathcal{H}}$ is the frequency-independent Hamiltonian with matrix given by
\begin{equation}
\hat{\mathcal{H}}=\hbar\left[\begin{array}{cccccccc}
\omega_{\tilde{\nu}}-\iu \Gamma & \cdots & 0 & \cdots & V_{\tilde{\nu},n}(z_0) & \cdots & V_{\tilde{\nu},m}(z_0) & \cdots  \\
\vdots &  \ddots &\vdots & & \vdots  & & \vdots &\\
0 &\cdots & \omega_{\tilde{\mu}}-\iu \Gamma & \cdots & V_{\tilde{\mu},n}(z_0) & \cdots & V_{\tilde{\mu},m}(z_0) & \cdots\\
\vdots & & \vdots & \ddots & \vdots & & \vdots &\\
V^\dagger_{\tilde{\nu},n}(z_0) & \cdots &V^\dagger_{\tilde{\mu},n}(z_0) & \cdots & \omega_n-\iu \gamma_n & \cdots &0 &\cdots \\
\vdots & &\vdots & & \vdots&\ddots & \vdots &\\
V^\dagger_{\tilde{\nu},m}(z_0) & \cdots &V^\dagger_{\tilde{\mu},m}(z_0) & \cdots & 0 & \cdots & \omega_m-\iu \gamma_m & \cdots\\
\vdots & &\vdots & & \vdots& & \vdots &\ddots\\
\end{array}\right].
\label{eq:15}
\end{equation}
Here, we can replaced the pre-fractor $\omega$ in the second term in the LHS of Eq.~\eqref{eq:12} as $\omega\to\sqrt{\omega_{\tilde{\nu}}\omega_{\tilde{\mu}}}$.  We note that the matrix of $\hat{\mathcal{H}}$ represents the generalized Hopfield matrix for heterostructure metasurfaces~\cite{gerace2007quantum}, and $\operatorname{dim}(\hat{\mathcal{H}})=[\operatorname{dim}(n)+\operatorname{dim}(\tilde{\nu})] \times[\operatorname{dim}(n)+\operatorname{dim}(\tilde{\nu})]$, where $\operatorname{dim}(\tilde{\nu})=2\operatorname{dim}(\nu)$.

Exciton-polariton BICs in both the weak and strong coupling regimes are solutions of $\operatorname{det}|\hat{\mathcal{H}}-\omega\hat{\mathcal{I}}|=0$ above the light cone with zero radiative losses. Figure~\ref{fig:0} shows schematically the real part of exciton and photon frequencies in the k-vector space in the weak coupling regime. The excitonic dispersion shown with yellow solid line is slow $\omega_{\tilde{\nu}}\simeq\omega_0$ as follows from Eq.~\eqref{eq:3}. The photonic dispersion shown with blue solid and dashed lines for different photonic designs is fast on the same scale of k-vectors. The solution of the weak-coupling regime is achieved by intersection of exciton and photon curves, and the modified exciton wavefunction can be treated as dressed with photonic states~\cite{kibis2011dissipationless}. 

\subsection{Weak-coupling regime and Purcell factor for single exciton}

We assume we are interested in the exciton-polaritons formed by light-matter coupling of a selected exciton state of a patterned vdW layer. In realistic vdW materials and semiconductor quantum wells, the effect of kintetic dispersion term and potential $\hat{\mathcal{U}}(x,y)$ in Eq.~\eqref{eq:3} is very weak~\cite{gerace2007quantum}, thus we the exciton frequency close to the frequency of a 2D exciton in an unpatterned layer $\omega_0$  in the frequency range of interest. Then, we can drop index $\tilde{\nu}$ in the exciton wavefunction and polarization in Eq.~\eqref{eq:10}, expand $\mathbf{d}=-\iu a_{\rm B} \sqrt{\pi\hbar c\Gamma_0/(4\omega_0)}\mathbf{e}$, and write the nonlinear eigenmode equation as 
\begin{equation}
\omega=\omega_0-\iu\Gamma+{2\Gamma_0\sum_{\mathbf{g},\mathbf{g}^\prime}\Phi^*_{\mathbf{g}}\mathbf{e}\cdot\hat{\mathbf{G}}_{\mathbf{g},\mathbf{g}^\prime}(z_0,z_0;\omega)\cdot\mathbf{e}^*\Phi_{\mathbf{g}^\prime}},
\label{eq:16}
\end{equation}
where we omitted superscript $\rm (ph)$ for the sake of brevity.

Equation~\eqref{eq:16} can be solved in the weak-coupling regime assuming $\operatorname{Re}[\omega]\simeq \omega_0$ and $\Gamma_0,\Gamma\ll \omega_0$. In this approximation, the spontaneous emission rate $\tilde{\Gamma}_0 $ of quantized excitons in the photonic environment of metasurface can be calculated from Eq.~\eqref{eq:16} as $\tilde{\Gamma}_0 =-\operatorname{Im}[\omega]-\Gamma$,
\begin{equation}
\tilde{\Gamma}_0  =-2\Gamma_0\sum_{\mathbf{g},\mathbf{g}^\prime}\operatorname{Im}\left[\Phi^*_{\mathbf{g}}\mathbf{e}\cdot\hat{\mathbf{G}}_{\mathbf{g},\mathbf{g}^\prime}(z_0,z_0;\omega_0)\cdot\mathbf{e}^*\Phi_{\mathbf{g}^\prime}\right].
\label{eq:17}
\end{equation}

The Purcell factor can be written by definition as ${\rm F}({\mathbf{k}_{\rm B}})\equiv{\tilde{\Gamma}_{0}(\mathbf{k}_{\rm B})}/{{\Gamma}_{0}(\mathbf{k}_{\rm B})}$, 
where ${\Gamma}_{0}(\mathbf{k}_{\rm B})$ is the vacuum spontaneous emission rate of exciton in an unpatterned vdW layer with in-plane Bloch vector $\mathbf{k}_{\rm B}$. We can evaluate ${\Gamma}_{0}(\mathbf{k}_{\rm B})$ via Eq.~\eqref{eq:17} with the use of the vacuum GF $\hat{\mathbf{G}}^{(\rm bg)}_{\mathbf{g},\mathbf{g}^\prime}(\mathbf{k}_{\rm B},z,z^\prime;k)$ satisfying periodic boundary conditions~\cite{chew1999waves}
\begin{equation}
\hat{\mathbf{G}}^{(\rm bg)}_{\mathbf{g},\mathbf{g}^\prime}(\mathbf{k}_{\rm B},z,z^\prime;k)=k\delta_{\mathbf{g
},\mathbf{g}^\prime}\left[\frac{\delta(z-z^{\prime})}{k^2}\mathbf{e}_z\otimes\mathbf{e}_{z}  + \frac{1}{2\iu k_{z,\mathbf{K}}(k)}\sum_{\alpha={\rm s,p}}\mathbf{f}^{(\sigma)}_{\alpha,\mathbf{K}}(z;k)\otimes\mathbf{f}^{(\sigma)}_{\alpha,\mathbf{K}^\prime}(-z^\prime;k)\right],
\label{eq:18}    
\end{equation}
where $k=\omega/c$, $\mathbf{K}=\mathbf{k}_{\rm B}+\mathbf{g}$ is the total in-plane k-vector, $k_{z,\mathbf{K}}(k)=\operatorname{sign}(k)\sqrt{k^2-\mathbf{K}^2}$, $\sigma(z,z^\prime)=\operatorname{sign}{(z-z^\prime)}$, the term with $\delta(z-z^{\prime})$ denotes the contribution of static modes~\cite{neale2020resonant},  and $\mathbf{f}^{(\sigma)}_{\alpha,\mathbf{K}}(z;k)$ are solutions of $k^2\mathbf{f}^{(\sigma)}_{\alpha,\mathbf{K}}(z;k)-\nabla_\mathbf{K}\times\nabla_\mathbf{K}\times\mathbf{f}^{(\sigma)}_{\alpha,\mathbf{K}}(z;k)=0$ given by
\begin{equation}
\mathbf{f}^{(\sigma)}_{\alpha,\mathbf{K}}(z;k) = \eu^{\iu \sigma k_{z,\mathbf{K}}(k)z}\mathbf{e}^{(\sigma)}_{\alpha,\mathbf{K}}(k).
\label{eq:19}    
\end{equation}
Here, $\mathbf{e}^{(\sigma)}_{\alpha,\mathbf{K}}$ are the unit polarization vectors for s and p polarization,
\begin{equation}
\mathbf{e}_{\rm s,\mathbf{K}} = \frac{1}{|\mathbf{K}|}\left[\begin{array}{c}
-K_y\\
K_x\\
0
\end{array}\right],\quad \mathbf{e}^{(\sigma)}_{\rm p,\mathbf{K}}(k) = \frac{1}{k|\mathbf{K}|}\left[\begin{array}{c}
K_x k_{z,\mathbf{K}}(k)\\
K_y k_{z,\mathbf{K}}(k)\\
-\sigma \mathbf{K}^2
\end{array}\right].
\label{eq:20}    
\end{equation}
Then,  ${\Gamma}_0(\mathbf{k}_{\rm B})$ evaluated from Eqs.~(\ref{eq:17}, \ref{eq:18}, \ref{eq:19}, \ref{eq:20}) for $\Phi_{\mathbf{g}}=\delta_{\mathbf{g},0}$.
\begin{equation}
{\Gamma}_{0}(\mathbf{k}_{\rm B};k_0) =\Gamma_0 H_\theta(|k_0|-|\mathbf{k}_{\rm B}|)\frac{\left[1-|\mathbf{k}_{\rm B}\cdot\mathbf{e}|^2/k_0^2\right]}{\sqrt{1-\mathbf{k}_{\rm B}^2/k_0^2}}= \Gamma_0 \left[1-|\mathbf{k}_{\rm B}\cdot\mathbf{e}|^2/k_0^2\right]\pi\rho_{\mathbf{k}_{\rm B}}(k_0),
\label{eq:21}    
\end{equation}
where $k_0=\omega_0/c$, $H_\theta$ is the Heaviside theta function, and $\rho_{\mathbf{K}}(k)$ is the one-dimensional vacuum LDOS into a radiation channel at in-plane vector $\mathbf{K}$ with polarization $\rm s$ (or $\rm p$)~\cite{ochiai2001nearly}
\begin{equation}
\rho_{\mathbf{K}}(k) \equiv  \frac{H_\theta(|k|-|\mathbf{K}|)}{\pi\sqrt{1-\mathbf{K}^2/k^2}}.
\label{eq:22}    
\end{equation}
The total number of quantum radiation states in the given interval of frequencies assuming a quantization size of $L$ can be calculated as
\begin{equation}
N_{\mathbf{K}}(k) = L\int^k_0 {\rm d}k^\prime\ \rho_{\mathbf{K}}(k^\prime)=\frac{L}{\pi}{\sqrt{k^2-\mathbf{K}^2}}H_\theta(k-|\mathbf{K}|).
\end{equation}
\label{eq:23}    

A bound (non-propagating to the outside environment) state is formed for $k_0$ satisfying $\tilde{\Gamma}_{0}(\mathbf{k}_{\rm B})=0$. To become a BIC, the frequency of bound state $k_0$ should lie within the radiation continuum, $\rho_{\mathbf{k}_{\rm B}+\mathbf{g}}(k_0)\ne 0$ for at least one $\mathbf{g}$. Using Eq.~\eqref{eq:21} and definition of Purcell factor, we can write $\tilde{\Gamma}({\mathbf{k}_{\rm B}})\propto{\rm F}({\mathbf{k}_{\rm B}})\rho_{\mathbf{k}_{\rm B}}$. In the subwavelength regime $k_{\rm BIC} p<2\pi$, the BIC frequency $k_{\rm BIC}$ can be obtained from the equation
\begin{equation}
\begin{cases}
\rm F({\mathbf{k}_{\rm B}};k_{\rm BIC})=0,\\
\rho_{\mathbf{k}_{\rm B}}(k_{\rm BIC})\ne 0.
\end{cases}
\label{eq:24}
\end{equation}
This equation is equivalent to Eq.~(3) of the main text.

\subsection{Excitonic BIC condition}

In this section, we find the expression for $k_{\rm BIC}$ in the weak coupling regime depending on the metasurface and exciton parameters. We start with the case of heterostructures with "weak" periodic patterning that can be treated within the perturbation theory and next derive the BIC condition in the general case. We consider the subwavelength regime, $k_0 p<2\pi$, with a single open radiation channel $\mathbf{g}=\mathbf{0}$ and normal incidence, $\mathbf{k}_{\rm B}=\mathbf{0}$. In this case, the Purcell factor of a 2D exciton in the heterostructure metasurface can be written from Eq.~\eqref{eq:17} as
\begin{equation}
{\rm F}(z_0;k_0)=\frac{\tilde{\Gamma}_0(z_0;k_0)}{\Gamma_{0}} =-2\sum_{\mathbf{g},\mathbf{g}^\prime}\operatorname{Im}\left[\Phi^*_{\mathbf{g}}\mathbf{e}\cdot\hat{\mathbf{G}}_{\mathbf{g},\mathbf{g}^\prime}(z_0,z_0;k_0)\cdot\mathbf{e}^*\Phi_{\mathbf{g}^\prime}\right].
\label{eq:25}   
\end{equation}

\subsubsection{Perturbation theory.} 
\label{sec:pert}
For the sake of clarity and brevity, we only consider (i) vdW layer coordinate in the center of metasurface $z_0=0$, (ii) s-polarized excitons at normal incidence $\mathbf{e}=\mathbf{e}_y$.

% Similarly, we can split the exciton Hamiltonian for the center-of-mass envelope wavefunction $\hat{\mathcal{H}}^{\rm (exc,env)}$ into the kinetic energy, averaged potential $\hat{\mathcal{U}}_{\mathbf{0}}$, and a periodic potential $\hat{\mathcal{U}}(x,y)$. In the Fourier domain, it can be written as  
% \begin{equation}
% \hat{\mathcal{H}}^{\rm (exc,env)}_{\mathbf{g},\mathbf{g}^{\prime}}=\delta_{\mathbf{g},\mathbf{g}^\prime}\left(\hbar \omega_0 + \frac{\hbar^2|\mathbf{g}|^2}{2 m_{\rm (exc)}} \delta_{\mathbf{g}, \mathbf{g}^{\prime}}+\hat{\mathcal{U}}_{\mathbf{0}}\right)+ \hat{\mathcal{U}}_{\mathbf{g}-\mathbf{g}^{\prime}},
% \end{equation}
% We only consider a real-valued potential with an even-order symmetry with respect to $x$ and $y$ that can be written as $\hat{\mathcal{U}}_{{g}_{x}-{g}_{y}}=\hat{\mathcal{U}}_{-{g}_{x}-{g}_{y}}=\hat{\mathcal{U}}_{{g}_{x}+{g}_{y}}=\hat{\mathcal{U}}_{-{g}_{x}+{g}_{y}}$.
% \begin{equation}
% \Delta \Phi_{g}=-\delta \hat{\mathcal{U}}_{g_{1,0}}\left(\delta_{g,g_{1,0}}+\delta_{g,g_{0,1}}+\delta_{g,g_{-1,0}}+\delta_{g,g_{0,-1}}\right),
% \label{eq:deltaphi}
% \end{equation}
% where $\delta \hat{\mathcal{U}}_g=\delta_{\mathbf{g\ne0}} \hat{\mathcal{U}}_g/({\hbar^2|g|^2}/{2 m_{\rm (exc)}})$ is a small perturbation parameter $\delta \hat{\mathcal{U}}_g\ll 1$.

We focus on the fundamental exciton state with the unperturbed wavefunction $\Phi^{(\rm wg)}_{\mathbf{g}}=\delta_{g,\mathbf{0}}$ and energy $\hbar\omega_0$. The perturbed wavefunction can be written as $\Phi_{\mathbf{g}}=\Phi^{(\rm wg)}_{\mathbf{g}}+\delta \Phi_{\mathbf{g}}$, where $\delta \Phi_{\mathbf{0}}=0$ and $|\delta \Phi_{\mathbf{g}\ne\mathbf{0}}|\ll 1$. Therefore, the Purcell factor expression in Eq.~\eqref{eq:25} can be simplified as
\begin{equation}
{\rm F}(0;k_0)\simeq-2\operatorname{Im}\left[\mathbf{e}_y\cdot\hat{\mathbf{G}}_{\mathbf{0},\mathbf{0}}(0,0;k_0)\cdot\mathbf{e}_y\right].
\label{eq:26}
\end{equation}

We assume the periodic resonant structure is formed by weakly perturbing the effective planar slab with a periodic perturbation. We can split the permittivity function of the periodic resonant structure $\varepsilon(\mathbf{r})$ into planar $\varepsilon(z)$, averaged perturbation $\varepsilon_{\mathbf{0}}(z)=\iint_{0}^p ({\rm d}x{\rm d}y/p^2)\ \varepsilon(\mathbf{r})$, and a periodic perturbation $\delta\varepsilon(\mathbf{r})$, where $p$ is the heterostructure period in both directions. In the reciprocal space, $\varepsilon_{\mathbf{g}}(z)$ can be written as 
\begin{equation}
\varepsilon_{\mathbf{g}}(z)=\overline{\varepsilon}(z)\left[\delta_{\mathbf{g},\mathbf{0}} + \delta\varepsilon_{\mathbf{g}}(z)\right].
\label{eq:35}
\end{equation}
where $\overline{\varepsilon}(z)=\varepsilon(z)+ \varepsilon_{\mathbf{0}}(z)$ is
the permittivity of an effective planar waveguide. For metasurface composed of a rectangular pattern of square block of width $W$, we can calculate it as $\overline{\varepsilon}=1+(\varepsilon-1) W^2/p^2$. We assume $\delta\varepsilon_{\mathbf{0}}(z)=0$ and the lossless material, so that $\delta\varepsilon_{\mathbf{g}}(z)=\delta\varepsilon^*_{-\mathbf{g}}(z)$.

The $\mathbf{g}=\mathbf{g}^\prime=\mathbf{0}$ Fourier component of the resonant GF of the periodic structure $\hat{\mathbf{G}}_{\mathbf{g},\mathbf{g}^\prime}(z,z^\prime;k)$ projected on $y$ axis at $z,z^\prime=0$ as $\left[\hat{\mathbf{G}}_{\mathbf{0},\mathbf{0}}(0,0;k)\right]_{y,y}$ can be expressed via the Dyson equation
\begin{equation}
\left[\hat{\mathbf{G}}_{\mathbf{0},\mathbf{0}}(0,0;k)\right]_{y,y}=\left[\hat{\mathbf{G}}^{(\rm wg)}_{\mathbf{0},\mathbf{0}}(0,0;k)\right]_{y,y}-k\int\limits_{-\infty}^\infty {\rm d}z \ \overline{\varepsilon}(z) \left[\hat{\mathbf{G}}^{(\rm wg)}_{\mathbf{0},\mathbf{0}}(0,z;k)\right]_{y,y}\sum_{\mathbf{g}}\delta\varepsilon_{-\mathbf{g}}(z)\left[\hat{\mathbf{G}}_{\mathbf{g},\mathbf{0}}(z,0;k)\right]_{y,y}.  
\label{eq:Dyson}
\end{equation}
Here, $\hat{\mathbf{G}}^{(\rm wg)}_{\mathbf{g},\mathbf{g}^\prime}(z,z^\prime;k)$ is the resonant GF of an effective waveguide with averaged permittivity $\overline{\varepsilon}(z)$. For $|z|,|z^\prime|<a$, the slab GF can be written in the form of Eq.~\eqref{eq:18} as~\cite{chew1999waves}
\begin{equation}
\hat{\mathbf{G}}^{(\rm wg)}_{\mathbf{g},\mathbf{g}^\prime}(z,z^\prime;k)=k\delta_{\mathbf{g
},\mathbf{g}^\prime}\left[\frac{\delta(z-z^{\prime})}{q^2(k)}\mathbf{e}_z\otimes\mathbf{e}_{z}  + \frac{1}{2\iu q_{z,\mathbf{g}}(k)}\sum_{\alpha={\rm s,p}}\frac{\mathbf{h}^{(\sigma)}_{\alpha,\mathbf{g}}(z;k)\otimes\mathbf{h}^{(\sigma)}_{\alpha,\mathbf{g}^\prime}(-z^\prime;k)}{\left(1-{r}_{\alpha,\mathbf{g}}^2(k)\mathrm{e}^{4\iu q_{z,\mathbf{g}}(k)a}\right)}\right],
\label{eq:003wg} 
\end{equation}
where $q(k)=\sqrt{\overline{\varepsilon}}k$, $q_{z,\mathbf{g}}(k)=\operatorname{sign}(k)\sqrt{q^2(k)-\mathbf{g}^2}$, $r_{\alpha,\mathbf{g}}(k)$ are Fresnel reflection coefficients at the core and cladding interface 
\begin{equation}
{r}_{{\rm s},\mathbf{g}}(k)=\frac{q_{z,\mathbf{g}}(k)-k_{z,\mathbf{g}}(k)}{q_{z,\mathbf{g}}(k)+k_{z,\mathbf{g}}(k)},\quad  {r}_{{\rm p},\mathbf{g}}(k)=\frac{q_{z,\mathbf{g}}(k)-\overline{\varepsilon}k_{z,\mathbf{g}}(k)}{q_{z,\mathbf{g}}(k)+\overline{\varepsilon}k_{z,\mathbf{g}}(k)},
\label{eq:rs}
\end{equation}
and and $\mathbf{h}^{(\sigma)}_{\alpha,\mathbf{g}}(z;k)$ are solutions of $q^2(k)\mathbf{h}^{(\sigma)}_{\alpha,\mathbf{g}}(z;k)-\nabla_\mathbf{g}\times\nabla_\mathbf{g}\times\mathbf{h}^{(\sigma)}_{\alpha,\mathbf{g}}(z;k)=0$ given by
\begin{equation}
\mathbf{h}^{(\sigma)}_{\alpha,\mathbf{g}}(z;k)\equiv\mathbf{f}^{(\sigma)}_{\alpha,\mathbf{g}}(z;q)+{r}_{\alpha,\mathbf{g}}(k)\mathrm{e}^{2\iu q_{z,\mathbf{g}}(k)a}\mathbf{f}^{(-\sigma)}_{\alpha,\mathbf{g}}(z;q).
\end{equation}

Assuming a uniform $\delta\varepsilon_{\mathbf{g}}(z)$ that is localized within $|z|\le a$, where $2a$ is the metasurface thickness, we can solve Eq.~\eqref{eq:Dyson} within the perturbation theory as $\hat{\mathbf{G}}_{\mathbf{g},\mathbf{g}^\prime}(z,z^\prime;k)=\hat{\mathbf{G}}^{(\rm wg)}_{\mathbf{g},\mathbf{g}^\prime}(z,z^\prime;k)+\Delta\hat{\mathbf{G}}_{\mathbf{g},\mathbf{g}^\prime}(z,z^\prime;k)$. Since $\delta\varepsilon_{\mathbf{0}}=0$, the lowest non-zero perturbation order is second,
\begin{equation}
\left[\Delta\hat{\mathbf{G}}_{\mathbf{0},\mathbf{0}}(0,0;k)\right]_{y,y}=k^2\overline{\varepsilon}^2\sum_\mathbf{g}|\delta\varepsilon_{\mathbf{g}}|^2\int\limits_{-a}^a {\rm d}z\int\limits_{-a}^a {\rm d}z^\prime\ \left[\hat{\mathbf{G}}^{(\rm wg)}_{\mathbf{0},\mathbf{0}}(0,z;k)\right]_{y,y}\left[\hat{\mathbf{G}}^{(\rm wg)}_{\mathbf{g},\mathbf{g}}(z,z^\prime;k)\right]_{y,y}\left[\hat{\mathbf{G}}^{(\rm wg)}_{\mathbf{0},\mathbf{0}}(z^\prime,0;k)\right]_{y,y}.
\label{eq:Dyson1}
\end{equation}

The Purcell factor can be written from Eq.~\eqref{eq:26} as
\begin{equation}
{\rm F}(0,\omega_0)={\rm F}^{(\rm wg)}(\omega_0)  + \Delta {\rm F}(\omega_0)+\sum_\mathbf{g}O(|\delta\Phi_\mathbf{g}\delta\varepsilon_{\mathbf{g}}|,|\delta\varepsilon_{\mathbf{g}}|^4,\ldots)
\label{eq:totFwo}
\end{equation}
where $|\delta\Phi_\mathbf{g}|\ll|\delta\varepsilon_{\mathbf{g}}|\ll 1$, and
\begin{align}
&{\rm F}^{(\rm wg)}(k)=-2\operatorname{Im}\left[\mathbf{e}_y\cdot\hat{\mathbf{G}}^{\rm (wg)}_{\mathbf{0},\mathbf{0}}(0,0;k)\cdot\mathbf{e}_y\right],\label{eq:wgF}\\
&\Delta {\rm F}(k)=-2\operatorname{Im}\left[\mathbf{e}_y\cdot\Delta\hat{\mathbf{G}}_{\mathbf{0},\mathbf{0}}(0,0;k)\cdot\mathbf{e}_y\right],\label{eq:deltaF}
\end{align}
where ${\rm F}^{(\rm wg)}(k)$ is the Purcell factor of an effective waveguide with $\overline{\varepsilon}$. 

Equation~\eqref{eq:wgF} can be evaluated by using Eq.~\eqref{eq:003wg}
\begin{equation}
{\rm F}^{(\rm wg)}(k)=\frac{1}{\left[\cos^2(q a)+\overline{\varepsilon}\sin^2(q a)\right]}.
\end{equation}
Equation~\eqref{eq:deltaF} can be further simplified by using Eq.~\eqref{eq:Dyson1} and the Mittag-Leffler expansion $\hat{\mathbf{G}}^{(\rm wg)}_{\mathbf{g},\mathbf{g}}(z,z^\prime;k)=\sum_m{\mathbf{E}_{m,\mathbf{g}}(z)\otimes\mathbf{E}_{m,-\mathbf{g}}(z^\prime)}/{(k-k_m)}$, where $k_m=(\omega_m -\iu\gamma_m)/c$ are complex frequencies. For the sake of simplicity, we only consider QNMs even with respect to $z\to -z$, so that $\Delta {\rm F}(k)$ in Eq.~\eqref{eq:deltaF} transforms into
\begin{equation}
\Delta {\rm F}(k)=-2q^2(k)\left[{\rm F}^{(\rm wg)}(k)\right]^2\sum_\mathbf{g}\left|\delta\varepsilon_{\mathbf{g}}\right|^2\operatorname{Im}\left[\sum_m\frac{A^2_{m,\mathbf{g}}(k)}{(k-k_m)}\right],
\label{eq:deltaF1}
\end{equation}
where $A_{m,-\mathbf{g}}(k)=A_{m,\mathbf{g}}(k)$ is an auxiliary function
\begin{equation}
A_{m,\mathbf{g}}(k)=   \frac{E_{m,\mathbf{g}}\left[q \cos(qa)-q\left(\cos(q_{z,\mathbf{g},m}a)+\iu \sqrt{\overline{\varepsilon}} \cos(qa)\right)+\iu \sqrt{\overline{\varepsilon}}q_{z,\mathbf{g},m}\sin(q_{z,\mathbf{g},m}a)\right]\left[\cos(qa)+\iu\sqrt{\overline{\varepsilon}}\sin(qa)\right]}{\left(q^2-q^2_{z,\mathbf{g},m}\right)},
\label{eq:AA}
\end{equation}
where $q_{z,\mathbf{g},m}=q_{z,\mathbf{g}}(k_m)$ and we used $\mathbf{E}_{m,\mathbf{g}}(z)=\mathbf{e}_{y} E_{m,\mathbf{g}}\cos{(q_{z,\mathbf{g},m}z)}$ with the normalization constant $E_{m,\mathbf{g}}$ given by~\cite{neale2020resonant}
\begin{equation}
E_{m,\mathbf{g}} = \left[2\overline{\varepsilon}a+\frac{2\mathbf{g}^2}{k_m^2\sqrt{\mathbf{g}^2-k_m^2}}\right]^{-1/2},
\end{equation}
and the resonant frequency $k_m$ defined as solution of
\begin{equation}
{r}_{{\rm s},\mathbf{g}}(k_m)=\mathrm{e}^{-2\iu q_{z,\mathbf{g},m}a}.
\end{equation}

The parameter $|\Delta\varepsilon_{\mathbf{g}}|/\overline{\varepsilon}$ in Eq.~\eqref{eq:deltaF1} is small, thus we consider $\omega_0$ close to a resonant frequency of the photonic GF $\hat{\mathbf{G}}^{(\rm wg)}_{\mathbf{g},\mathbf{g}}(0,0;k)$ for a specific $\mathbf{g}$ to compensate for the small numerator. In particular, we consider the sub-diffraction regime $|\mathbf{g}_{1,0}|>k_m >|\mathbf{g}_{1,0}|/\sqrt{\overline{\varepsilon}}$ and choose a real-valued pole $\hat{\mathbf{G}}^{(\rm wg)}_{\mathbf{g}_{1,0},\mathbf{g}_{1,0}}(0,0;k)$ corresponding to two degenerate s-polarized guided modes with frequency $k_m=\omega_m/c$, where $\mathbf{g}_{l,s}=2\pi /p [l\mathbf{e}_x\ s\mathbf{e}_y]^\mathsf{T}$. For our choice of s-polarized guided modes, the summation over $\mathbf{g}_{l,s}$ in Eq.~\eqref{eq:deltaF1} includes only $l=\pm 1$ $s=0$, so the BIC condition ${\Delta {\rm F}(\omega_{\rm BIC})}=-{{\rm F}^{(\rm wg)}(\omega_m)}$ can be written as 
\begin{equation}
{\omega_{\rm BIC}}/{\omega_m}=8\left|\delta\varepsilon_{\mathbf{g}_{1,0}}\right|^2\overline{\varepsilon}k_m{\rm F}^{(\rm wg)}(k_m)\operatorname{Re}\left[A_{m,\mathbf{g}_{1,0}}(k_m)\right]\operatorname{Im}\left[A_{m,\mathbf{g}_{1,0}}(k_m)\right],
\label{eq:BICcondition0}
\end{equation}
valid for asymptotically for $|\delta\Phi_\mathbf{g}|\ll|\delta\varepsilon_{\mathbf{g}}|\to 0$. We note that $\Delta {\rm F}(k_m)\to \infty$, which indicates limited applicability of the perturbation approach in the current form, as it does not account for the strong coupling between two degenerate guided modes with the frequency $k_m$ propagating to the right and left with the reciprocal vector $\mathbf{g}_{1,0}$.

We can evaluate $|\delta\varepsilon_{\mathbf{g}_{1,0}}|$ for centered square meta-atoms of width $W=p(1-\delta p)$ in a rectangular grid with period $p$ as
$\delta\varepsilon_{\mathbf{g}_{l,s}}=\delta_{l,s\ne0}(-1)^{l+s}(1/\varepsilon-1)\sin(\pi l\delta p)/(\pi l)\sin(\pi s\delta p)/(\pi s)$. We can consider the width of the air gaps to be small, $\delta p\ll 1$, so that $|\delta\varepsilon_{\mathbf{g}_{l,s}}|^2\simeq \delta_{l,s\ne0} (1/\varepsilon-1)^2\delta p^4$.

\subsubsection{General case}

In this subsection, we derive a general condition of excitonic BIC for an arbitrary periodic heterostructure. We first obtain the pole expansion of the Purcell factor into QNM contributions by substituting the GF pole expansion Eq.~\eqref{eq:11} into Eq.~\eqref{eq:25}, 
\begin{equation}
{\rm F}(z_0;\omega_0)=\frac{1}{\Gamma_0}\operatorname{Im}\left[-\sum_{n} \frac{V_{n}(z_0)V_{n}^\dagger(z_0)}{(\omega_0-\omega_n+\iu\gamma_n)}\right],
\label{eq:purex}
\end{equation}
where the exciton-photon coupling amplitudes are defined as in Eq.~\eqref{eq:13}. We note that Eq.~\eqref{eq:purex} is a known result (see Ref.~\cite{muljarov2016exact}) adapted for a periodic system with a patterned 2D exciton layer.

We consider the exciton frequency $\omega_0$ in the vicinity of a high-Q guided-mode resonance (GMR) with the complex frequency $\omega_n-\iu\gamma_n$. Then, we separate the term corresponding to the GMR in Eq.~\eqref{eq:purex},
\begin{equation}
{\rm F}(z_0;\omega_0)= {\rm F}^{(\rm nres)}(z_0;\omega_0)-\frac{1}{\Gamma_0}\operatorname{Im}\left[\frac{V_{n}(z_0)V^\dagger_{n}(z_0)}{(\omega_0-\omega_n+\iu\gamma_n)}\right],
\label{eq:FPnew3}
\end{equation}
where ${\rm F}^{(\rm nres)}(z_0;\omega_0)$ is a slow function of $\omega_0$ describing the contribution of non-resonant modes, that include the Fabry-Perot (FP) modes, other non-resonant GMR modes, and Rayleign anomaly contribution at the diffraction threshold frequencies, that can be discretized into so-called cut modes~\cite{neale2020resonant}.

Equation~\eqref{eq:FPnew3} can be re-written in the form of a generalized Fano formula, using mathematical approach developed in Ref.~\cite{bogdanov2019bound},
\begin{equation}
{\rm F}(z_0;\omega_0)=\left[{\rm F}^{(\rm nres)}(z_0;\omega_0)-{\rm F}_n^{(\rm env)}(z_0)\right]+\frac{\left[q_n(z_0)+\Delta \omega_{0,n}\right]^2}{1+(\Delta \omega_{0,n})^2}{\rm F}_n^{(\rm env)}(z_0).
\label{eq:FPnew4}
\end{equation}
Here, $\Delta \omega_{0,n}\equiv(\omega_0-\omega_n)/\gamma_n$, $q_n$ is the Fano parameter, ${\rm F}_n^{(\rm env)}$ is the smooth envelope, that are given by
\begin{equation}
\begin{aligned}
&q_n(z_0)=-\cot{\left\{\frac{\arg{\left[V_{n}(z_0)V^\dagger_{n}(z_0)\right]}}{2}\right\}},\\
&{\rm F}_n^{(\rm env)}(z_0)=\frac{\left|V_{n}(z_0)V^\dagger_{n}(z_0)\right|}{\gamma_n\Gamma_0[1+q_n^2(z_0)]}.
\end{aligned}
\label{eq:QF}
\end{equation}
The minimal value of ${\rm F}(z_0;\omega_0)$ in Eq.~\eqref{eq:FPnew4} is achieved at $\Delta \omega_{0,n}=-q_n(z_0)$ and is equal to ${\rm F}^{(\rm nres)}(z_0;\omega_0)-{\rm F}_n^{(\rm env)}(z_0)$. Therefore, the BIC condition can be written as
\begin{equation}
\begin{aligned}
& \omega_{\rm BIC}=\omega_n-q_n(z_0)\gamma_n,\\
&{\rm F}^{(\rm nres)}(z_0;\omega_{\rm BIC})={\rm F}_n^{(\rm env)}(z_0).
\end{aligned}    
\label{eq:BICcondition2}
\end{equation}

It is an open question, whether both conditions in Eq.~\eqref{eq:BICcondition2} can be fulfilled exactly. We next study the validity of Eq.~\eqref{eq:BICcondition2} in the asymptotic limit of small perturbations and weakly coupled modes. As in Sec.~\ref{sec:pert}, we only consider (i) vdW layer coordinate in the center of metasurface $z_0=0$, (ii) s-polarized excitons at normal incidence $\mathbf{e}=\mathbf{e}_y$, (iii) eigenfrequency $\omega_n-\iu \gamma_n$ to correspond to a bright GMR formed via the strong coupling between two degenerate s-polarized guided modes with frequency $\omega_m$ propagating to the right and left with the reciprocal vectors $\pm\mathbf{g}_{1,0}$.

Within the perturbation theory, we can find the the GMR mode frequency 
\begin{equation}
\begin{aligned}
&\omega_n/\omega_m=1+\left|\delta\varepsilon_{\mathbf{g}_{1,0}}\right|^2\left(\chi_{\mathbf{g}_{1,0}}(q_m)+4\overline{\varepsilon}k_m{\rm F}^{(\rm wg)}(k_m)\operatorname{Re}\left[A_{m,\mathbf{g}_{1,0}}(k_m)\right]\operatorname{Im}\left[A_{m,\mathbf{g}_{1,0}}(k_m)\right]\right),\\
&\gamma_n/\omega_m = 4\left|\delta\varepsilon_{\mathbf{g}_{1,0}}\right|^2\overline{\varepsilon}k_m{\rm F}^{(\rm wg)}(k_m)\operatorname{Im}^2\left[A_{m,\mathbf{g}_{1,0}}(k_m)\right],
\end{aligned}
\label{eq:GMR2}
\end{equation}
and the electric field at $z=0$
\begin{equation}
\mathbf{E}_{n,\mathbf{g}}(0)=\left[-\delta_{\mathbf{g},\mathbf{0}}\sqrt{2} q_m|\delta\varepsilon_{\mathbf{g}_{1,0}}|{\rm F}^{(\rm wg)}(k_m)A_{m,\mathbf{g}_{1,0}}(k_m)+\delta_{\mathbf{g},\mathbf{g}_{1,0}}\frac{\delta\varepsilon_{\mathbf{g}_{1,0}}}{\sqrt{2}|\delta\varepsilon_{\mathbf{g}_{1,0}}|}+\delta_{\mathbf{g},-\mathbf{g}_{1,0}}\frac{\delta\varepsilon^*_{\mathbf{g}_{1,0}}}{\sqrt{2}|\delta\varepsilon_{\mathbf{g}_{1,0}}|}
\right]\mathbf{E}_{m,\mathbf{g}_{1,0}}(0).
\label{eq:GMR3}    
\end{equation}
Here, $\chi_{\mathbf{g}_{1,0}}(q_m)$ is an auxiliary function given by
\begin{equation}
\chi_{\mathbf{g}_{1,0}}=\frac{2\overline{\varepsilon}q^2_mE^2_{m,{\mathbf{g}_{1,0}}}}{\mathbf{g}_{1,0}^2}\left[a+\frac{\sin(2q_{z,\mathbf{g}_{1,0},m}a)}{2q_{z,\mathbf{g}_{1,0},m}}-\frac{2\left[q_m \cos(q_{z,\mathbf{g}_{1,0},m}a)\sin(q_ma)-q_{z,\mathbf{g}_{1,0},m} \sin(q_{z,\mathbf{g}_{1,0},m}a)\cos(q_ma)\right]}{\mathbf{g}_{1,0}^2} \right],   
\end{equation}
where $q_m=q(k_m)$.

The coupling coefficients are given by Eq.~\eqref{eq:13}, $V_{n}(z)=\sqrt{2c\Gamma_0}\sum_\mathbf{g}\Phi^*_{\mathbf{g}}\mathbf{e}_{y}\cdot\mathbf{E}_{n,\mathbf{g}}(z)$ and $V^\dagger_{n}(z)=\sqrt{2c\Gamma_0}\sum_\mathbf{g}\Phi_{\mathbf{g}}\mathbf{e}_{y}\cdot\mathbf{E}_{n,-\mathbf{g}}(z)$. As in Sec.~\ref{sec:pert}, the perturbed wavefunction can be written as $\Phi_{\mathbf{g}}=\delta_{\mathbf{g},\mathbf{0}}+\delta \Phi_{\mathbf{g}}$, where $\delta \Phi_{\mathbf{0}}=0$ and $|\delta \Phi_{\mathbf{g}\ne\mathbf{0}}|\ll 1$. Then, $V_{n}(z_0)\simeq V^\dagger_{n}(z_0)\simeq\sqrt{2c\Gamma_0}\mathbf{e}_{y}\cdot\mathbf{E}_{n,\mathbf{0}}(z_0)$ in Eq.~\eqref{eq:BICcondition2} can be calculated at $z=z_0=0$ via Eq.~\eqref{eq:GMR3} as
\begin{equation}
V_{n}(0) \simeq -2|\delta\varepsilon_{\mathbf{g}_{1,0}}|\sqrt{c\Gamma_0}q_m{\rm F}^{(\rm wg)}(k_m)A_{m,\mathbf{g}_{1,0}}(k_m).
\label{eq:VV}
\end{equation}
We can substitute Eq.~\eqref{eq:VV} into Eq.~\eqref{eq:QF} can see that 
\begin{equation}
\begin{aligned}
&q_n(0)=-\cot{\left\{{\arg{\left[V_{n}(0)\right]}}\right\}}=-\frac{\operatorname{Re}[A_{m,\mathbf{g}_{1,0}}(k_m)]}{\operatorname{Im}[A_{m,\mathbf{g}_{1,0}}(k_m)]},\\
&{\rm F}^{(\rm env)}_n(0)=\frac{\operatorname{Im}^2[V_{n}(0)]}{\gamma_n\Gamma_0}={\rm F}^{(\rm wg)}(k_m).
\end{aligned}   
\label{eq:QF1}
\end{equation}
The BIC condition can be written by substituting Eqs.~(\ref{eq:GMR2}, \ref{eq:QF1}) into  Eq.~\eqref{eq:BICcondition2} as
\begin{equation}
\begin{aligned}
& {\omega_{\rm BIC}}/{\omega_m}=1+\left|\delta\varepsilon_{\mathbf{g}_{1,0}}\right|^2\left(\chi_{\mathbf{g}_{1,0}}(q_m)+8\overline{\varepsilon}k_m{\rm F}^{(\rm wg)}(k_m)\operatorname{Re}\left[A_{m,\mathbf{g}_{1,0}}(k_m)\right]\operatorname{Im}\left[A_{m,\mathbf{g}_{1,0}}(k_m)\right]\right),\\
&{\rm F}^{(\rm nres)}(0;\omega_m)\simeq{\rm F}^{(\rm wg)}(\omega_m).
\end{aligned}    
\label{eq:BICcondition3}
\end{equation}
The first of Eq.~\eqref{eq:BICcondition3} has the form of Eq.~\eqref{eq:BICcondition0} with an additional perturbation term $\chi_{\mathbf{g}_{1,0}}(q_m)$ accounting for the strong coupling of two guided modes in this model. The second of Eq.~\eqref{eq:BICcondition3} requires ${\rm F}^{(\rm nres)}(z_0,\omega_m)\simeq {\rm F}^{(\rm wg)}(z_0,\omega_m)$ that can be realized for high-index materials and $\omega_m$ away from the diffraction threshold frequency $2\pi c/p$. We study the functional behavior of ${\rm F}^{(\rm wg)}(z_0,\omega_0)$ in the next section~\ref{sec:planar}.

\section{Effective dielectric slab with a 2D vdW layer}
\label{sec:planar}

In this section, we analyze the Purcell factor contribution by an effective dielectric slab ${\rm F}^{(\rm wg)}(z_0,\omega_0)$ in detail. We consider a planar dielectric slab with effective permittivity $\overline{\varepsilon}(z)$ reconstructed from the metasurface permittivity profile as
\begin{equation}
\overline{\varepsilon}(z)=\varepsilon(z)+ \iint_{0}^p ({\rm d}x{\rm d}y)/p^2\ \Delta\varepsilon(\mathbf{r}).    
\end{equation}
For the sake of brevity, we only consider s (TE) polarization.

\subsection{Purcell factor at oblique incidence for TE polarization}

% \begin{figure}
%     \centering
%     \includegraphics[width=0.8\linewidth]{xxx.pdf}
%     \caption{.}
%     \label{fig:S1}
% \end{figure}

\subsubsection{Definitions}

The transversal component of the electric-electric GF inside the slab for $|z|,|z^\prime|\le a$ for TE polarization at the Bloch wavevector $\mathbf{k}_B$ can be evaluated as ${\rm G}^{(\rm wg)}(s,z, z^\prime;k)=\mathbf{e}_{\rm s,\mathbf{k}_{\rm B}}\cdot\hat{\mathbf{G}}_{\mathbf{0},\mathbf{0}}^{\rm (wg)}(\mathbf{k}_{\rm B},z, z^\prime;k)\cdot\mathbf{e}_{\rm s,\mathbf{k}_{\rm B}}$ using Eq.~\eqref{eq:003wg}, where $s=|\mathbf{k}_B|/k$. Then, we can write the Maxwell equation for ${\rm G}^{(\rm wg)}(z, z^\prime;k)$ as
\begin{equation}
\left[\partial^2_z+k^2 \left(\overline{\varepsilon}(z) -s^2\right)\right]{\rm G}^{(\rm wg)}(s,z, z^{\prime};k)=k\delta(z-z^{\prime}),
\end{equation}
with the outgoing B.C. at the outer interface
\begin{equation}
\left.\left[\partial_z \mp \iu k_z(k)\right] {\rm G}^{(\rm wg)}(s,z, z^{\prime};k)\right|_{z= \pm a}=0,
\end{equation}
and continuity B.C. for ${\rm G}^{(\rm wg)}(s,z, z^{\prime};k)$ and $\partial_z{\rm G}^{(\rm wg)}(s,z, z^{\prime};k)$ for $z$ on the inner and outer interfaces.

The analytic form of ${\rm G}^{(\rm wg)}(z, z^\prime;k)$ can be written as
\begin{equation}
{\rm G}^{(\rm wg)}(s,z, z^\prime;k)=\frac{k\left[\eu^{\iu q_z |z-z^\prime|} +r^{2}\eu^{4\iu q_za}\eu^{-\iu q_z |z-z^\prime|} + 2r\eu^{2\iu q_za}\cos{(q_z z+q_zz^\prime)}\right]}{2 \iu q_z(k)\left[1-\eu^{4\iu q_za}r^2\right]}.
\label{eq:S1}
\end{equation}
Here, $q_z(k)=k\sqrt{\overline{\varepsilon}-s^2}$, $k=\omega/c$, $s=\sin\theta$ and $\theta\in[0,\pi/2]$ is the angle of incidence. The internal Fresnel reflection amplitude $r$ for s polarization  is given by Eq.~\eqref{eq:rs}
\begin{equation}
r(s)=\frac{\sqrt{\overline{\varepsilon}-s^2}-\sqrt{1-s^2}}{\sqrt{\overline{\varepsilon}-s^2}+\sqrt{1-s^2}}.
\label{eq:S2}
\end{equation}
where $k_z(k)=k\sqrt{1-s^2}$. The free-space background GF can be obtained from Eq.~\eqref{eq:S1} in the limit $\overline{\varepsilon}\to 1$
\begin{equation}
{\rm G}^{\rm (bg)}(s,z, z^\prime;k)=\frac{k\eu^{\iu k_z |z-z^\prime|}}{2 \iu k_z(k)}.
\label{eq:S3}
\end{equation}
We note that ${\rm G}^{\rm (bg)}(s,z, z^\prime;k)$ is polarization independent.

\subsubsection{Purcell factor evaluation and analysis}

The Purcell factor for $s$-polarized excitons can be evaluated via Eq.~\eqref{eq:003wg}. We can evaluate the imaginary part of ${\rm G}^{(\rm wg)}$ and ${\rm G}^{\rm (bg)}$ at $z=z^\prime=z_0$ and $k=k_0$ by using Eqs.~(\ref{eq:S1},~\ref{eq:S3}),
\begin{equation}
\begin{aligned}
&\operatorname{Im}\left[{\rm G}^{(\rm wg)}(s,z_0, z_0;k_0)\right] =-\frac{k\left(1-r\right)^2\left[1+r^{2}  +2r\cos(2 q_z a)\cos(2 q_zz_0)\right]}{2q_z(k_0)\left[1+r^{4}  -2r^{2}\cos(4 q_z a)\right]},\\
&\operatorname{Im}\left[{\rm G}^{\rm (bg)}(s,z_0, z_0;k_0)\right] =  -\frac{k}{2 k_z(k_0)}.
\end{aligned}    
\end{equation}
Thus, the Purcell factor is
\begin{equation}
{\rm F}^{(\rm wg)}(s,z_0;k_0) = \left(1-r\right)^2\frac{\left[1+r^{2}  +2r\cos(2 q_z a)\cos(2 q_zz_0)\right]}{\left[(1+r^{2})^2  -4r^{2}\cos^2(2 q_z a)\right]}.
\label{eq:S01010}
\end{equation}
where we used $(1+r)k_z(k_0)=(1-r)q_z(k_0)$ from Eq.~\eqref{eq:S2}. We can introduce dimensionless $\tilde{z}_0=z_0/a$ and $\tilde{\omega}_0(s)=2 q_z a=\omega_0\sqrt{1-(s^2/\overline{\varepsilon})}/\omega_{\rm ph}$, $\omega_{\rm ph} = \pi c/(2\sqrt{\overline{\varepsilon}}a)$ and re-write Eq.~\eqref{eq:S01010} as
\begin{equation}
{\rm F}^{(\rm wg)}(s,\tilde{z}_0;\tilde{\omega}_0) = \left(1-r\right)^2\frac{\left[1+r^{2}  +2r\cos(\pi \tilde{\omega}_0)\cos(\pi \tilde{\omega}_0\tilde{z}_0)\right]}{\left[(1+r^{2})^2  -4r^{2}\cos^2(\pi \tilde{\omega}_0)\right]}.
\label{eq:S0101}
\end{equation}

We next find the local extrema of ${\rm F}^{(\rm wg)}$ in Eq.~\eqref{eq:S0101} as a function of $\tilde{z}_0$ and $k_0$. The set of equations $\partial_{\tilde{z}_0}{\rm F}^{(\rm wg)}(\tilde{z}_0;\tilde{\omega}_0) =0$ and $\partial_{\tilde{\omega}_0}{\rm F}^{(\rm wg)}(\tilde{z}_0;\tilde{\omega}_0) =0$ for $0\le s<1$ and $|\tilde{z}_0|\le 1$ can be written as
\begin{equation}
\begin{cases}
\sin(\pi \tilde{\omega}_0\tilde{z}_0)\cos(\pi \tilde{\omega}_0)=0,\\
\sin(\pi \tilde{\omega}_0)\frac{\left[2\left(n^2-1\right)\left(n^2+1-2 s^2\right) \cos(\pi \tilde{\omega}_0)+\left((n^2+1-2s^2)^2+\left(n^2-1\right)^2 \cos^2(\pi \tilde{\omega}_0)\right) \cos(\pi \tilde{\omega}_0\tilde{z}_0)\right]}{\left[(n^2+1-2s^2)-\left(n^2-1\right) \cos(\pi \tilde{\omega}_0)\right]\left[(n^2+1-2s^2)+\left(n^2-1\right) \cos(\pi \tilde{\omega}_0)\right]}=0.
\label{eq:S4}
\end{cases}
\end{equation}
Equations~\eqref{eq:S4} have as two independent sets of solutions given by
\begin{align}
&\begin{cases}
\sin(\pi \tilde{\omega}_0\tilde{z}_0)=0,\\
\sin(\pi \tilde{\omega}_0)=0,
\end{cases}\label{eq:S51}\\
&\begin{cases}
\cos(\pi \tilde{\omega}_0\tilde{z}_0)=0,\\
\cos(\pi \tilde{\omega}_0)=0.
\end{cases}
\label{eq:S52}
\end{align}
In particular, Eq.~\eqref{eq:S51} solutions are
\begin{equation}
\begin{cases}
\tilde{\omega}_{0}=\omega_0/\omega_{\rm ph}(s)=m, &\ m=1,2,\ldots,\\
\tilde{z}_0=l/m, &\ l=-m,\ldots m.
\end{cases}
\label{eq:solcomp}
\end{equation}
Equations~\eqref{eq:solcomp} correspond to Eqs.~(5,6) of the main text

We next analyze the sign of $\mathcal{D}= (\partial^2_{\tilde{z}_0}{\rm F}^{(\rm wg)})(\partial^2_{\tilde{\omega}_0}{\rm F}^{(\rm wg)}) -(\partial_{\tilde{z}_0,\tilde{\omega}_0}^2{\rm F}^{(\rm wg)})^2$ and $\partial^2_{\tilde{z}_0}{\rm F}^{(\rm wg)}$ at $\tilde{z}_0$ and $\tilde{\omega}_0$ corresponding to solutions of Eqs.~(\ref{eq:S51},~\ref{eq:S52}):
\begin{itemize}
    \item  First pair of sets of equations derived from Eq.~\eqref{eq:S51} describes even solutions
\begin{equation}
\begin{cases}
\sin(\pi \tilde{\omega}_0\tilde{z}_0/2)=0,\\
\sin(\pi \tilde{\omega}_0/2)=0,
\end{cases}\quad \text{and odd solutions} \quad \begin{cases}
\cos(\pi \tilde{\omega}_0\tilde{z}_0/2)=0,\\
\cos(\pi \tilde{\omega}_0/2)=0,
\end{cases}
\label{eq:sol1}
\end{equation}
that correspond to ${\rm F}^{(\rm wg)}_{\rm max}=1$. It is a {\bf maximum  point} because of $\mathcal{D}=\pi^2(\overline{\varepsilon}-1)^2\tilde{\omega}_0^2/[(\overline{\varepsilon}-s^2)(1-s^2)]>0$ and $\partial^2_{\tilde{z}_0}{\rm F}^{(\rm wg)}=-\pi^2(\overline{\varepsilon}-1) \tilde{\omega}_0^2/(2\overline{\varepsilon}-2s^2)<0$. The even solutions of Eqs.~\eqref{eq:sol1} can be written as
\begin{equation}
\begin{cases}
\tilde{\omega}_{0}(s)=2m, &\ m=1,2,\ldots,\\
\tilde{z}_0=l/m, &\ l=-m,\ldots m,
\end{cases}
\quad \text{and solutions as} \quad \begin{cases}
\tilde{\omega}_{0}(s)=2m+1,&\ m=1,2,\ldots,\\
\tilde{z}_0=(2l+1)/(2m+1), &\ l=-m,\ldots m;
\end{cases}
\label{eq:S061}
\end{equation}

\item Second pair of sets of equations derived from Eq.~\eqref{eq:S51} describes even solutions
\begin{equation}
\begin{cases}
\sin(\pi \tilde{\omega}_0\tilde{z}_0/2)=0,\\
\cos(\pi \tilde{\omega}_0/2)=0,
\end{cases}\quad \text{and odd solutions as} \quad \begin{cases}
\cos(\pi \tilde{\omega}_0\tilde{z}_0/2)=0,\\
\sin(\pi \tilde{\omega}_0/2)=0,
\end{cases}
\label{eq:sol2}
\end{equation}
that correspond to ${\rm F}^{(\rm wg)}_{\rm min}=(1-s^2)/(\overline{\varepsilon}-s^2)$. It is a {\bf minimum point} because of $\mathcal{D}=\pi^2(\overline{\varepsilon}-1)^2(1-s^2)\tilde{\omega}_0^2/(\overline{\varepsilon}-s^2)^3>0$ and $\partial^2_{\tilde{z}_0}{\rm F}^{(\rm wg)}=\pi^2(\overline{\varepsilon}-1)\tilde{\omega}_0^2/(2\overline{\varepsilon}-2s^2)>0$. The even solutions of Eqs.~\eqref{eq:sol2} can be written as
\begin{equation}
\begin{cases}
\tilde{\omega}_{0}(s)=2m, &\ m=1,2,\ldots,\\
\tilde{z}_0=(2l+1)/(2m), &\ l=-m,\ldots m-1,
\end{cases}
\ \text{and odd solutions} \ \begin{cases}
\tilde{\omega}_{0}(s)=2m+1,&\ m=1,2,\ldots,\\
\tilde{z}_0=2l/(2m+1), &\ l=-m,\ldots m.
\end{cases}
\label{eq:S6}
\end{equation}

The minimum value of ${\rm F}^{(\rm wg)}_{\rm min}$ achieved at Eqs.~\eqref{eq:S6} is a function of the angle of incidence $\theta$ and permittivity, 
\begin{equation}
{\rm F}^{(\rm wg)}_{\rm min}=\frac{\cos^2\theta}{(\overline{\varepsilon}-\sin^2\theta)}. 
\end{equation}
The background radiative decay rate changes with the angle of incidence as $
-2\operatorname{Im}\left[{\rm G}^{\rm (bg)}(s,\tilde{z}_0, \tilde{z}_0;\tilde{\omega}_0)\right]\Gamma_0={\Gamma_0}/{\cos\theta}$,
which coincides with the earlier results~\cite{ivchenko2005optical}. Therefore, the minimal dressed exciton radiative decay rate achieved at Eqs~\eqref{eq:S6} is
\begin{equation}
\tilde{\Gamma}_{0,\rm min}(\theta)= -2\operatorname{Im}\left[{\rm G}^{\rm (bg)}(s,\tilde{z}_0, \tilde{z}_0;\tilde{\omega}_0)\right]\Gamma_0{\rm F}^{(\rm wg)}_{\rm min}=  \frac{\cos\theta}{(\overline{\varepsilon}-\sin^2\theta)}\Gamma_0;
\end{equation}

\item Third set of equations given by Eq.~\eqref{eq:S52}
\begin{equation}
\begin{cases}
\cos(\pi \tilde{\omega}_0\tilde{z}_0)=0,\\
\cos(\pi \tilde{\omega}_0)=0.
\end{cases}
\label{eq:sol4}
\end{equation}
corresponds to ${\rm F}^{(\rm wg)}_{\rm sdl}=2(1-s^2)/(1+\overline{\varepsilon}-2s^2)$. It is a {\bf saddle point} because of $\mathcal{D}=-16\pi^2(\overline{\varepsilon}-1)^2(1-s^2)^2\tilde{\omega}_0^2/(1+\overline{\varepsilon}-2s^2)^4<0$. The solutions of Eqs.~\eqref{eq:sol4} can be written as
\begin{equation}
\begin{cases}
\tilde{\omega}_{0}(s)=m+1/2, &\ m=1,2,\ldots,\\
\tilde{z}_0=(2l+1)/(2m+1), &\ l=-m,\ldots m.
\end{cases}
\label{eq:S761}
\end{equation}

We note that for $\tilde{z}_0=\pm 1$ at the slab's surface, the saddle point given by Eqs.~\eqref{eq:S761} is a {\bf minimum point} for the function of one variable ${\rm F}^{(\rm wg)}(\pm1;\tilde{\omega}_0)$, with the minimal value of
\begin{equation}
{\rm F}^{(\rm wg)}_{\rm min}=\frac{2\cos^2(\theta)}{\overline{\varepsilon}+1-2\sin^2(\theta)}.    
\end{equation}
\end{itemize}

\subsubsection{Mode analysis}
%% FIGURE S1
%%%%%%%%%%%%%%%%%%%%%%%%%%%%%%%%%%%%%%%
\begin{figure*}[t]
\includegraphics[width=0.7\linewidth]{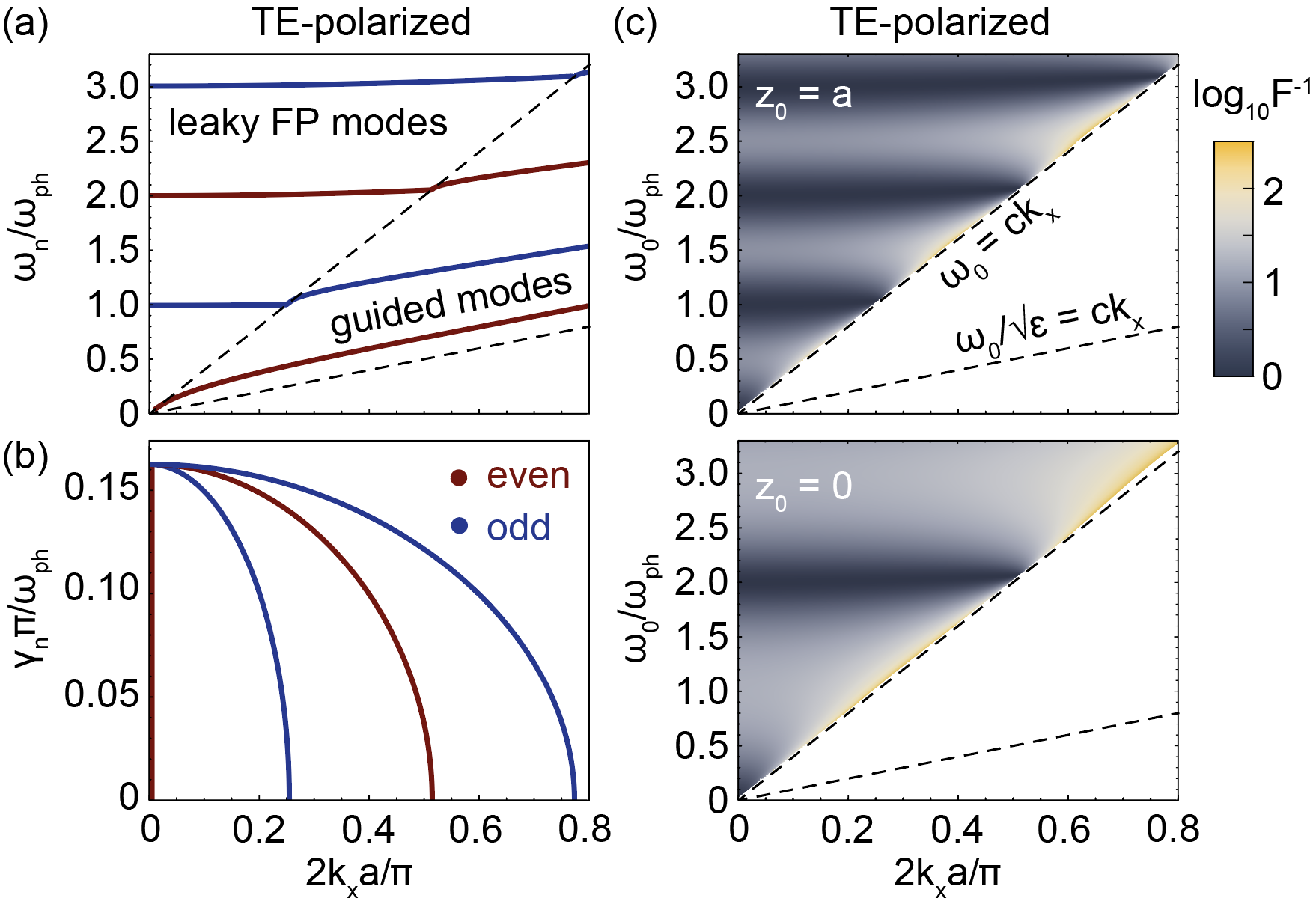}% Here is how to import EPS art
\caption{\label{fig:S1} {\bf Mode spectrum and Purcell factor for dielectric slab at oblique incidence for TE polarization}. (a,b) Dispersion diagram $\omega_n/\omega_{\rm ph}$ (a) and radiative decay rate $\gamma_n\pi/\omega_{\rm ph}$ (b) vs. $2k_xa/\pi$ for FP QNMs given by Eqs.~(\ref{eq:modes},~\ref{eq:modesS}). Here, $\omega_{\rm ph} = \pi c/(2\sqrt{\overline{\varepsilon}}a)$. Red and blue line color shows even and odd parity with respect to $z\to -z$ mirror symmetry, respectively. (c) Logarithmic scale of ${\rm F}^{-1}$ vs. $\tilde{\omega}_0$ and  $2k_xa/\pi$ for $ z_0=a$ (top) and ${z}_0=0$ (bottom) given by Eq.~\eqref{eq:S0101}.}
\end{figure*}
%%%%%%%%%%%%%%%%%%%%%%%%%%%%%%%%%%%%%%%
In this subsection, we analyze how the maxima and minima conditions in Eqs.~(\ref{eq:S061},\ref{eq:S6}) correspond to the Fabry--P\'{e}rot (FP) quasi-normal modes (QNMs) of the slab. The only non-zero component of electric field of TE-polarized FP QNMs with complex frequencies $k_m=(\omega_m-\iu\gamma_m)/c$ satisfies~\cite{neale2020resonant}
\begin{equation}
\left[\partial^2_z+\overline{\varepsilon}(z)k_m^2-k_x^2\right]E_m(z;k_x)=0,
\end{equation}
with the outgoing B.C. at the outer interface
\begin{equation}
\left.\left[\partial_z \mp \iu \sqrt{k_m^2-k_x^2}\right] E_m(z;k_x)\right|_{z= \pm a}=0,
\end{equation}
and continuity B.C. for ${\rm G}(z, z^{\prime})E_m(z;k_x)$ and $\partial_zE_m(z;k_x)$ for $z$ on the inner and outer interfaces.

We consider only TE-polarized FP QNMs with $\omega_m>0$. These mode frequencies are given by~\cite{neale2020resonant}
\begin{equation}
\begin{aligned}
&\omega_m=\operatorname{Re}\left[\sqrt{\left(m -\iu\frac{\ln{(r_m^{-1})}}{\pi}\right)^2+\left(\frac{2k_x a}{\pi}\right)^2}\right]\omega_{\rm ph},\quad m=1,2,\ldots,\\
&\gamma_m=-\operatorname{Im}\left[\sqrt{\left(m -\iu\frac{\ln{(r_m^{-1})}}{\pi}\right)^2+\left(\frac{2k_x a}{\pi}\right)^2}\right]\omega_{\rm ph}.
\end{aligned}
\label{eq:modes}
\end{equation}
Here, $\omega_{\rm ph} = \pi c/(2\sqrt{\overline{\varepsilon}}a)$ is the same as before, and $r_m=r(k_m)$ is the TE reflection coefficient in Eq.~\eqref{eq:S2} at the complex QNM frequency, 
\begin{equation}
r_m=\frac{\sqrt{\overline{\varepsilon}k_m^2-k_x^2}-\sqrt{k_m^2-k_x^2}}{\sqrt{\overline{\varepsilon}k_m^2-k_x^2}+\sqrt{k_m^2-k_x^2}}.
\end{equation}
For $\gamma_m\ll\omega_m$, Eq.~\eqref{eq:modes} can be simplified as
\begin{equation}
\begin{aligned}
&\omega_m \simeq \sqrt{m^2+\left(\frac{2k_x a}{\pi}\right)^2}\omega_{\rm ph}, \quad m=1,2,\ldots,\\
&\gamma_m \simeq\frac{\omega_{\rm ph}}{\pi}\sqrt{1+\left(\frac{2k_x a}{\pi n}\right)^2}\ln{(r_m^{-1})} ,\\
&r_m\simeq \frac{\sqrt{\overline{\varepsilon}\omega_m^2/c^2-k_x^2}-\sqrt{\omega_m^2/c^2-k_x^2}}{\sqrt{\overline{\varepsilon}\omega_m^2/c^2 -k_x^2}+\sqrt{\omega_m^2/c^2-k_x^2}}.
\end{aligned}
\label{eq:modesS}
\end{equation}
The first equation for $\omega_m$ coincides with the first of Eqs.~\eqref{eq:solcomp}. Comparison of FP QNM spectra and Purcell factor spectra is shown in Fig.~\ref{fig:S1}.

The magnitude of the QNM electric field is given by~\cite{neale2020resonant}
\begin{equation}
|E_m(\tilde{z};k_x)|^2=\frac{\cos[\pi m(\tilde{z}-1)]+\cosh[\tilde{z}\ln{(r_m^{-1})}]}{4\left[\overline{\varepsilon} a+\frac{\iu k_x^2}{k_m^2\sqrt{k_n^2-k_x^2}}\right]}. 
\label{eq:Emodes}
\end{equation}
The field extrema are given by $\partial_{\tilde{z}}|E_m(\tilde{z};k_x)|^2=0$, that can be evaluated from Eq.~\eqref{eq:Emodes} as solutions of
\begin{equation}
\sin[\pi m(\tilde{z}-1)]=\frac{\ln{(r_m^{-1})}}{\pi m}\sinh[\tilde{z}\ln{(r_m^{-1})}]\simeq\frac{4\tilde{z}(1-s^2)}{\pi m(\sqrt{\overline{\varepsilon}-s^2}-\sqrt{1-s^2})^2}\simeq \frac{1}{m \overline{\varepsilon} }\simeq0.
\end{equation}
As a result, the field extrema are approximately
\begin{equation}
\tilde{z}\simeq l/m,\quad l=-m,\ldots m,
\end{equation}
which coincides with the second of Eqs.~\eqref{eq:solcomp}.

\subsection{Effective number of interacting Fabry--P\'{e}rot modes}
We consider the weak-coupling regime, thus we can treat Eq.~\eqref{eq:purex} within the perturbation theory. At normal incidence $k_x=0$, the amplitudes of coupling elements are given by Eq.~\eqref{eq:13} that is simplified to
\begin{equation}
V_{m}(z)=V^\dagger_{m}(z)=\sqrt{2c\Gamma_0}E_{m}(z).
\label{eq:SS13}
\end{equation}

The canonical perturbation strength is the ratio $\eta_m$ between $|V_m|$ and the complex frequency mismatch between the exciton and the $n$-th QNM, defined as
\begin{equation}
\eta_m(\tilde{z}_0; \tilde{\omega}_0)=\frac{|V_m(\tilde{z}_0)|}{\sqrt{(\omega_0-\omega_{n})^2+(\gamma-\Gamma)^2}}\simeq\frac{|V_m(\tilde{z}_0)|}{\sqrt{(\tilde{\omega}_0-m)^2\omega_{\rm ph}^2+\gamma^2}},
\end{equation}
where we assumed $\Gamma\ll \gamma$. We consider that the exciton is resonant with some $l$-th QNM $\tilde{\omega}_0=u$. Then,
\begin{equation}
\eta_m=\frac{\sqrt{\Gamma_0/\gamma}\sqrt{(\cos[\pi m(\tilde{z}_0-1)]+\cosh[2\tilde{z}_0/\sqrt{\overline{\varepsilon}}])/2}}{\sqrt{\pi^2\overline{\varepsilon}(u-m)^2/4+1}},    
\end{equation}
where we used $\omega_{\rm ph}=\pi\sqrt{\overline{\varepsilon}}\gamma/2$ and $\gamma=c/(\overline{\varepsilon} a)$. Then, we can compare $\eta_u$ with $\eta_{m\ne u}$ for different values of $\tilde{z}_0$.
\begin{itemize}
\item  $\tilde{z}_0=0$, then the perturbation strength parameter $\eta_m$ is simplified as
\begin{equation}
\eta_m =  \frac{\sqrt{\Gamma_0/\gamma}\sqrt{[1+(-1)^m]/2}}{\sqrt{\pi^2\overline{\varepsilon}(u/2-m/2)^2+1}}.
\end{equation}
For odd $u$ we get,
\begin{equation}
\begin{aligned}
&\eta_u=  0, \\ 
&\eta_{m\ne u}\simeq  \frac{2\sqrt{\Gamma_0/(\overline{\varepsilon}\gamma)}}{\pi|n-u|}, \quad    \text{even}\ m,\\
&\eta_{m\ne u}=  0, \quad    \text{odd}\ m. 
\end{aligned}    
\end{equation}
The Purcell factor is minimal in this regime, and many even-order QNMs with  $m\ne u$ are required to describe the suppression effect, because of $\eta_{m\ne u}>\eta_{u}=0$. For even $u$ we get,
\begin{equation}
\begin{aligned}
&\eta_u=  \sqrt{\Gamma_0/\gamma},\\ 
&\eta_{m\ne u}\simeq  \frac{2\sqrt{\Gamma_0/(\overline{\varepsilon}\gamma)}}{\pi|m-u|}, \quad    \text{even}\ m,\\
&\eta_{m\ne u}=  0, \quad    \text{odd}\ m. 
\end{aligned}    
\end{equation}
The Purcell factor is maximal ${\rm F}^{(\rm wg)}=1$ in this regime, and a single $u$-th QNM is sufficient, because of $\eta_{m\ne u}\ll\eta_{u}$.

\item $\tilde{z}_0 = l/m\ne 0,\pm 1$ and $(l-m)$ is odd. Then,  then $\eta_m$ is simplified as 
\begin{equation}
\eta_m=\frac{\sqrt{\Gamma_0/\gamma}\sqrt{((-1)^{l-m}+\cosh[2\tilde{z}_0/\sqrt{\overline{\varepsilon}}])/2}}{\sqrt{\pi^2\overline{\varepsilon}(u-m)^2/4+1}}=\frac{\sqrt{\Gamma_0/(\overline{\varepsilon}\gamma)}\tilde{z}_0}{\sqrt{\pi^2\overline{\varepsilon}(u-m)^2/4+1}}.   
\end{equation}
Then, 
\begin{equation}
\begin{aligned}
&\eta_u\simeq\sqrt{\Gamma_0/(\overline{\varepsilon}\gamma)}\tilde{z}_0,\\
&\eta_{m\ne u}\simeq\frac{2\sqrt{\Gamma_0/(\overline{\varepsilon}\gamma)}\tilde{z}_0}{\pi|m-u|}\simeq\frac{\eta_u}{|m-u|}.
\end{aligned}   
\end{equation}
The Purcell factor is minimal in this regime,  and many QNMs with  $m\ne u$ are required to describe the suppression effect, because of $\eta_{m\ne u}\simeq\eta_{u}$.

\item  $\tilde{z}_0= \pm 1$ and $\tilde{\omega}_0=u$. Then $\eta_m$ is simplified as
\begin{equation}
\eta_m \simeq  \frac{\sqrt{\Gamma_0/\gamma}}{\sqrt{\pi^2\overline{\varepsilon}(u-m)^2/4+1}}.
\end{equation}
Then,
\begin{equation}
\begin{aligned}
&\eta_u\simeq\sqrt{\Gamma_0/\gamma},\\
&\eta_{m\ne u}\simeq\frac{2\sqrt{\Gamma_0/(\overline{\varepsilon}\gamma)}}{\pi|m-u|}\simeq \frac{\eta_l}{\sqrt{\overline{\varepsilon}}|m-u|}.
\end{aligned}   
\end{equation}
The Purcell factor is maximal in this regime, and a single $u$-th QNM is sufficient, because of $\eta_{m\ne u}\ll\eta_{u}$.

\item  $\tilde{z}_0= \pm 1$ and $\tilde{\omega}_0=u+1/2$. Then $\eta_m$ is simplified as
\begin{equation}
\eta_m \simeq  \frac{\sqrt{\Gamma_0/\gamma}}{\sqrt{\pi^2\overline{\varepsilon}(u+1/2-m)^2/4+1}}\simeq\frac{2\sqrt{\Gamma_0/(\overline{\varepsilon}\gamma)}}{\pi|m-u-1/2|}.
\end{equation}
The Purcell factor is minimal in this regime, and many QNMs with $m/u\simeq 1$ are required, because of comparable $\eta_{m}$.

\end{itemize}
%% FIGURE S2
%%%%%%%%%%%%%%%%%%%%%%%%%%%%%%%%%%%%%%%
\begin{figure}[t]
\includegraphics[width=0.95\linewidth]{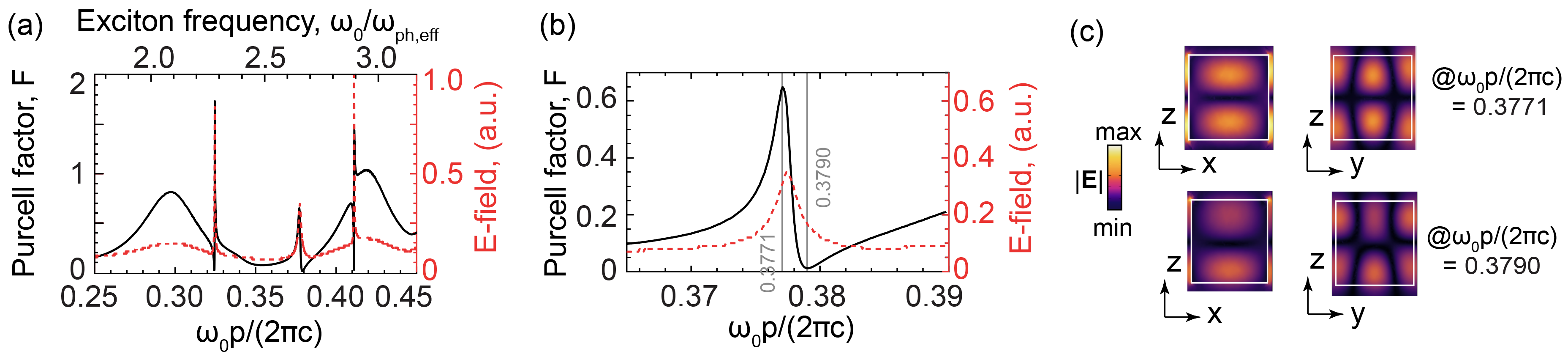}% Here is how to import EPS art
\caption{\label{fig:S2} {\bf Purcell factor and field enhancement in the quasi-BIC regime}. (a,b) Purcell factor (black) and averaged electric field norm (red) at the top surface of the metasurface in the broad range (a) and short range (b) of exciton frequencies. (c) induced electric field profiles at the maximum of Purcell factor ($0.3771$) and minimum ($0.3790$).}
\end{figure}
%%%%%%%%%%%%%%%%%%%%%%%%%%%%%%%%%%%%%%%

\section{Additional results for heterostructure metasurfaces}

We consider a metasurface composed of a square lattice of period $p$ containing dielectric rectangular meta-atom bars with permittivity $\varepsilon$, thickness $2a$ and width $W=0.9p$, see Fig.~4(a) of the main text. The FP mode contribution can be evaluated via an effective slab mode with $\overline{\varepsilon}=1+(\varepsilon-1)W^2/p^2 = 0.19+0.81\varepsilon$~\cite{andreani2006photonic}. We only consider normal incidence below.

\subsection{Field enhancement: numerical analysis}

Figure~\ref{fig:S2} shows the comparison of the near-field and far-field characteristics. The metasurface parameters are as in Fig. 4 of the main text. Panel (a) shows the comparison of Purcell factor [same as Fig.4b of the main text] and the electric field averaged over the vdW layer plane. One can see that the increase of the Purcell factor coincides with the enhancement of the electric field, but the suppression of the Purcell factor does not lead to the minimum of electric field. Panel (b) shows the zoom of the frequency range from $0.365$ to $0.390$ in dimensionless units. Similarly, the electric field is not minimal at the point of minimal Purcell factor at the frequency of $0.3790$. Panel (c) shows the electric field profiles at the frequencies of $0.3771$ and $0.3790$ that correspond to the Purcell factor maximum and minimum in panel (b), respectively. One can see that the field at the top surface (white lines outline the metasurface boundaries) is not minimal and barely changes from $0.3771$ to $0.3790$, which confirms that there is no field value influence on the Purcell factor value

%% FIGURE S3
%%%%%%%%%%%%%%%%%%%%%%%%%%%%%%%%%%%%%%%
\begin{figure}[t]
\includegraphics[width=0.75\linewidth]{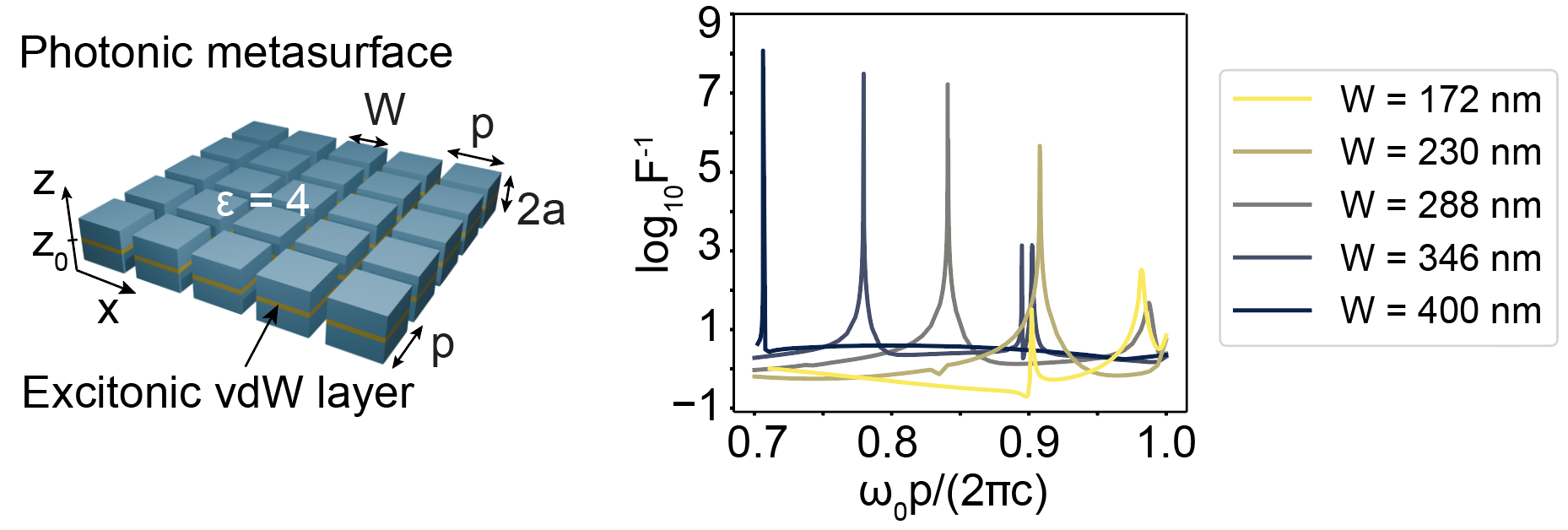}% Here is how to import EPS art
\caption{\label{fig:S3} {\bf Purcell factor calculations for low-contrast dielectric metasurface}. (left) Schematic of a heterostructure metasurface of thickness $2a=400$ nm, period $p=412$ nm, meta-atom width and length $W$, and permittivity $4$. (right) Inverse Purcell factor for the 2D vdW layer in the center of the metasurface at $z_0 = 0$, for $k_x = 0$ and TE polarization for various $W$ from $172$~nm to $400$~nm.}
\end{figure}
%%%%%%%%%%%%%%%%%%%%%%%%%%%%%%%%%%%%%%%

\subsection{Low-contrast heterostructure metasurface: numerical analysis}

To highlight the difference of excitonic BICs to other suppression mechanisms, we perform additional calculations for a metasurface composed of a low-index dielectric with permittivity of $4$, shown in Fig.~\ref{fig:S3}. The period is $p=412$~nm, and the square meta-atom width $W$ is varied from $172$~nm to $400$~nm, and the respective maximal inverse Purcell factor, corresponding to the excitonic quasi-BIC formation, changes from $550$ for $W = 172$~nm to $1.1\times 10^8$ for $W = 400$~nm. One can see robustness of the excitonic quasi-BIC with the geometrical parameter changes and increase of the Q factor for larger W that signifies asymptotic transition to the true excitonic BIC. 

\bibliography{bibliography}

\end{document}